\RequirePackage{lineno}
\documentclass[letterpaper,aps,prl,twocolumn,showpacs,superscriptaddress,groupedaddress,floatfix]{revtex4} 
\usepackage{graphicx}  
\usepackage{dcolumn}   
\usepackage{bm}        
\usepackage{amssymb}   

\newcommand{\invisible}[1]{}

\newcommand{\etadet}{\ensuremath{\eta_{\mathrm{det}}}}
\def\tanb  {\ensuremath{\mathrm{\tan \beta}}\xspace}

\RequirePackage{xspace}
\def\b     {\ensuremath{b}\xspace}
\def\bbar  {\ensuremath{\overline b}\xspace}

\newcommand{\gevcc}{\ensuremath{{\mathrm{\,Ge\kern -0.1em V\!/}c^2}}\xspace}
\newcommand{\gevc}{\ensuremath{{\mathrm{\,Ge\kern -0.1em V\!/}c}}\xspace}
\def\bc   {\begin{center}}
\def\ec   {\end{center}}
\newcommand{\jprlBase}       {Phys.\ Rev.\ Lett.\xspace}
\newcommand{\jprl}      [1]  {\jprlBase\ {\bf #1}}

\newcommand{\ifb}{\ensuremath{\mathrm{fb}^{-1}}}

\begin{document}
\hspace{5.2in} \mbox{Fermilab-Pub-12/385-E}
\title{Search for Neutral Higgs Bosons in Events with Multiple Bottom Quarks at the Tevatron}
\affiliation{LAFEX, Centro Brasileiro de Pesquisas F\'{i}sicas, Rio de Janeiro, Brazil}
\affiliation{Universidade do Estado do Rio de Janeiro, Rio de Janeiro, Brazil}
\affiliation{Universidade Federal do ABC, Santo Andr\'e, Brazil}
\affiliation{Institute of Particle Physics: McGill University, Montr\'{e}al, Qu\'{e}bec, Canada H3A~2T8; Simon Fraser University, Burnaby, British Columbia, Canada V5A~1S6; University of Toronto, Toronto, Ontario, Canada M5S~1A7; and TRIUMF, Vancouver, British Columbia, Canada V6T~2A3}
\affiliation{University of Science and Technology of China, Hefei, People's Republic of China}
\affiliation{Institute of Physics, Academia Sinica, Taipei, Taiwan 11529, Republic of China}
\affiliation{Universidad de los Andes, Bogot\'a, Colombia}
\affiliation{Charles University, Faculty of Mathematics and Physics, Center for Particle Physics, Prague, Czech Republic}
\affiliation{Czech Technical University in Prague, Prague, Czech Republic}
\affiliation{Center for Particle Physics, Institute of Physics, Academy of Sciences of the Czech Republic, Prague, Czech Republic}
\affiliation{Universidad San Francisco de Quito, Quito, Ecuador}
\affiliation{Division of High Energy Physics, Department of Physics, University of Helsinki and Helsinki Institute of Physics, FIN-00014, Helsinki, Finland}
\affiliation{LPC, Universit\'e Blaise Pascal, CNRS/IN2P3, Clermont, France}
\affiliation{LPSC, Universit\'e Joseph Fourier Grenoble 1, CNRS/IN2P3, Institut National Polytechnique de Grenoble, Grenoble, France}
\affiliation{CPPM, Aix-Marseille Universit\'e, CNRS/IN2P3, Marseille, France}
\affiliation{LAL, Universit\'e Paris-Sud, CNRS/IN2P3, Orsay, France}
\affiliation{LPNHE, Universit\'es Paris VI and VII, CNRS/IN2P3, Paris, France}
\affiliation{CEA, Irfu, SPP, Saclay, France}
\affiliation{IPHC, Universit\'e de Strasbourg, CNRS/IN2P3, Strasbourg, France}
\affiliation{IPNL, Universit\'e Lyon 1, CNRS/IN2P3, Villeurbanne, France and Universit\'e de Lyon, Lyon, France}
\affiliation{III. Physikalisches Institut A, RWTH Aachen University, Aachen, Germany}
\affiliation{Physikalisches Institut, Universit\"at Freiburg, Freiburg, Germany}
\affiliation{II. Physikalisches Institut, Georg-August-Universit\"at G\"ottingen, G\"ottingen, Germany}
\affiliation{Institut f\"{u}r Experimentelle Kernphysik, Karlsruhe Institute of Technology, D-76131 Karlsruhe, Germany}
\affiliation{Institut f\"ur Physik, Universit\"at Mainz, Mainz, Germany}
\affiliation{Ludwig-Maximilians-Universit\"at M\"unchen, M\"unchen, Germany}
\affiliation{Fachbereich Physik, Bergische Universit\"at Wuppertal, Wuppertal, Germany}
\affiliation{University of Athens, 157 71 Athens, Greece}
\affiliation{Panjab University, Chandigarh, India}
\affiliation{Delhi University, Delhi, India}
\affiliation{Tata Institute of Fundamental Research, Mumbai, India}
\affiliation{University College Dublin, Dublin, Ireland}
\affiliation{Istituto Nazionale di Fisica Nucleare Bologna, $^{\sharp{a}}$University of Bologna, I-40127 Bologna, Italy}
\affiliation{Laboratori Nazionali di Frascati, Istituto Nazionale di Fisica Nucleare, I-00044 Frascati, Italy}
\affiliation{Istituto Nazionale di Fisica Nucleare, Sezione di Padova-Trento, $^{\sharp{b}}$University of Padova, I-35131 Padova, Italy}
\affiliation{Istituto Nazionale di Fisica Nucleare Pisa, $^{\sharp{c}}$University of Pisa, $^{\sharp{d}}$University of Siena and $^{\sharp{e}}$Scuola Normale Superiore, I-56127 Pisa, Italy}
\affiliation{Istituto Nazionale di Fisica Nucleare, Sezione di Roma 1, $^{\sharp{f}}$Sapienza Universit\`{a} di Roma, I-00185 Roma, Italy}
\affiliation{Istituto Nazionale di Fisica Nucleare Trieste/Udine, I-34100 Trieste, $^{\sharp{g}}$University of Udine, I-33100 Udine, Italy}
\affiliation{Okayama University, Okayama 700-8530, Japan}
\affiliation{Osaka City University, Osaka 588, Japan}
\affiliation{Waseda University, Tokyo 169, Japan}
\affiliation{University of Tsukuba, Tsukuba, Ibaraki 305, Japan}
\affiliation{Center for High Energy Physics: Kyungpook National University, Daegu 702-701, Korea; Seoul National University, Seoul 151-742, Korea; Sungkyunkwan University, Suwon 440-746, Korea; Korea Institute of Science and Technology Information, Daejeon 305-806, Korea; Chonnam National University, Gwangju 500-757, Korea; Chonbuk National University, Jeonju 561-756, Korea}
\affiliation{Korea Detector Laboratory, Korea University, Seoul, Korea}
\affiliation{CINVESTAV, Mexico City, Mexico}
\affiliation{Nikhef, Science Park, Amsterdam, the Netherlands}
\affiliation{Radboud University Nijmegen, Nijmegen, the Netherlands}
\affiliation{Joint Institute for Nuclear Research, Dubna, Russia}
\affiliation{Institute for Theoretical and Experimental Physics, Moscow, Russia}
\affiliation{Moscow State University, Moscow, Russia}
\affiliation{Institute for High Energy Physics, Protvino, Russia}
\affiliation{Petersburg Nuclear Physics Institute, St. Petersburg, Russia}
\affiliation{Comenius University, 842 48 Bratislava, Slovakia; Institute of Experimental Physics, 040 01 Kosice, Slovakia}
\affiliation{Institut de Fisica d'Altes Energies, ICREA, Universitat Autonoma de Barcelona, E-08193, Bellaterra (Barcelona), Spain}
\affiliation{Instituci\'{o} Catalana de Recerca i Estudis Avan\c{c}ats (ICREA) and Institut de F\'{i}sica d'Altes Energies (IFAE), Barcelona, Spain}
\affiliation{Centro de Investigaciones Energeticas Medioambientales y Tecnologicas, E-28040 Madrid, Spain}
\affiliation{Instituto de Fisica de Cantabria, CSIC-University of Cantabria, 39005 Santander, Spain}
\affiliation{Uppsala University, Uppsala, Sweden}
\affiliation{University of Geneva, CH-1211 Geneva 4, Switzerland}
\affiliation{Glasgow University, Glasgow G12 8QQ, United Kingdom}
\affiliation{Lancaster University, Lancaster LA1 4YB, United Kingdom}
\affiliation{University of Liverpool, Liverpool L69 7ZE, United Kingdom}
\affiliation{Imperial College London, London SW7 2AZ, United Kingdom}
\affiliation{University College London, London WC1E 6BT, United Kingdom}
\affiliation{The University of Manchester, Manchester M13 9PL, United Kingdom}
\affiliation{University of Oxford, Oxford OX1 3RH, United Kingdom}
\affiliation{University of Arizona, Tucson, Arizona 85721, USA}
\affiliation{Ernest Orlando Lawrence Berkeley National Laboratory, Berkeley, California 94720, USA}
\affiliation{University of California, Davis, Davis, California 95616, USA}
\affiliation{University of California, Los Angeles, Los Angeles, California 90024, USA}
\affiliation{University of California Riverside, Riverside, California 92521, USA}
\affiliation{Yale University, New Haven, Connecticut 06520, USA}
\affiliation{University of Florida, Gainesville, Florida 32611, USA}
\affiliation{Florida State University, Tallahassee, Florida 32306, USA}
\affiliation{Argonne National Laboratory, Argonne, Illinois 60439, USA}
\affiliation{Fermi National Accelerator Laboratory, Batavia, Illinois 60510, USA}
\affiliation{Enrico Fermi Institute, University of Chicago, Chicago, Illinois 60637, USA}
\affiliation{University of Illinois at Chicago, Chicago, Illinois 60607, USA}
\affiliation{Northern Illinois University, DeKalb, Illinois 60115, USA}
\affiliation{Northwestern University, Evanston, Illinois 60208, USA}
\affiliation{University of Illinois, Urbana, Illinois 61801, USA}
\affiliation{Indiana University, Bloomington, Indiana 47405, USA}
\affiliation{Purdue University Calumet, Hammond, Indiana 46323, USA}
\affiliation{University of Notre Dame, Notre Dame, Indiana 46556, USA}
\affiliation{Purdue University, West Lafayette, Indiana 47907, USA}
\affiliation{Iowa State University, Ames, Iowa 50011, USA}
\affiliation{University of Kansas, Lawrence, Kansas 66045, USA}
\affiliation{Kansas State University, Manhattan, Kansas 66506, USA}
\affiliation{Louisiana Tech University, Ruston, Louisiana 71272, USA}
\affiliation{The Johns Hopkins University, Baltimore, Maryland 21218, USA}
\affiliation{Boston University, Boston, Massachusetts 02215, USA}
\affiliation{Northeastern University, Boston, Massachusetts 02115, USA}
\affiliation{Harvard University, Cambridge, Massachusetts 02138, USA}
\affiliation{Massachusetts Institute of Technology, Cambridge, Massachusetts 02139, USA}
\affiliation{Tufts University, Medford, Massachusetts 02155, USA}
\affiliation{University of Michigan, Ann Arbor, Michigan 48109, USA}
\affiliation{Wayne State University, Detroit, Michigan 48201, USA}
\affiliation{Michigan State University, East Lansing, Michigan 48824, USA}
\affiliation{University of Mississippi, University, Mississippi 38677, USA}
\affiliation{University of Nebraska, Lincoln, Nebraska 68588, USA}
\affiliation{Rutgers University, Piscataway, New Jersey 08855, USA}
\affiliation{Princeton University, Princeton, New Jersey 08544, USA}
\affiliation{University of New Mexico, Albuquerque, New Mexico 87131, USA}
\affiliation{State University of New York, Buffalo, New York 14260, USA}
\affiliation{The Rockefeller University, New York, New York 10065, USA}
\affiliation{University of Rochester, Rochester, New York 14627, USA}
\affiliation{State University of New York, Stony Brook, New York 11794, USA}
\affiliation{Brookhaven National Laboratory, Upton, New York 11973, USA}
\affiliation{Duke University, Durham, North Carolina 27708, USA}
\affiliation{The Ohio State University, Columbus, Ohio 43210, USA}
\affiliation{Langston University, Langston, Oklahoma 73050, USA}
\affiliation{University of Oklahoma, Norman, Oklahoma 73019, USA}
\affiliation{Oklahoma State University, Stillwater, Oklahoma 74078, USA}
\affiliation{University of Pennsylvania, Philadelphia, Pennsylvania 19104, USA}
\affiliation{Carnegie Mellon University, Pittsburgh, Pennsylvania 15213, USA}
\affiliation{University of Pittsburgh, Pittsburgh, Pennsylvania 15260, USA}
\affiliation{Brown University, Providence, Rhode Island 02912, USA}
\affiliation{University of Texas, Arlington, Texas 76019, USA}
\affiliation{Texas A\&M University, College Station, Texas 77843, USA}
\affiliation{Southern Methodist University, Dallas, Texas 75275, USA}
\affiliation{Rice University, Houston, Texas 77005, USA}
\affiliation{Baylor University, Waco, Texas 76798, USA}
\affiliation{University of Virginia, Charlottesville, Virginia 22904, USA}
\affiliation{University of Washington, Seattle, Washington 98195, USA}
\affiliation{University of Wisconsin, Madison, Wisconsin 53706, USA}
\author{T.~Aaltonen$^{\dag}$}~\affiliation{Division of High Energy Physics, Department of Physics, University of Helsinki and Helsinki Institute of Physics, FIN-00014, Helsinki, Finland}
\author{V.M.~Abazov$^{\ddag}$}~\affiliation{Joint Institute for Nuclear Research, Dubna, Russia}
\author{B.~Abbott$^{\ddag}$}~\affiliation{University of Oklahoma, Norman, Oklahoma 73019, USA}
\author{B.S.~Acharya$^{\ddag}$}~\affiliation{Tata Institute of Fundamental Research, Mumbai, India}
\author{M.~Adams$^{\ddag}$}~\affiliation{University of Illinois at Chicago, Chicago, Illinois 60607, USA}
\author{T.~Adams$^{\ddag}$}~\affiliation{Florida State University, Tallahassee, Florida 32306, USA}
\author{G.D.~Alexeev$^{\ddag}$}~\affiliation{Joint Institute for Nuclear Research, Dubna, Russia}
\author{G.~Alkhazov$^{\ddag}$}~\affiliation{Petersburg Nuclear Physics Institute, St. Petersburg, Russia}
\author{A.~Alton$^{\ddag a}$}~\affiliation{University of Michigan, Ann Arbor, Michigan 48109, USA}
\author{B.~\'{A}lvarez~Gonz\'{a}lez$^{\dag a}$}~\affiliation{Instituto de Fisica de Cantabria, CSIC-University of Cantabria, 39005 Santander, Spain}
\author{G.~Alverson$^{\ddag}$}~\affiliation{Northeastern University, Boston, Massachusetts 02115, USA}
\author{S.~Amerio$^{\dag}$}~\affiliation{Istituto Nazionale di Fisica Nucleare, Sezione di Padova-Trento, $^{\sharp{b}}$University of Padova, I-35131 Padova, Italy}
\author{D.~Amidei$^{\dag}$}~\affiliation{University of Michigan, Ann Arbor, Michigan 48109, USA}
\author{A.~Anastassov$^{\dag b}$}~\affiliation{Fermi National Accelerator Laboratory, Batavia, Illinois 60510, USA}
\author{A.~Annovi$^{\dag}$}~\affiliation{Laboratori Nazionali di Frascati, Istituto Nazionale di Fisica Nucleare, I-00044 Frascati, Italy}
\author{J.~Antos$^{\dag}$}~\affiliation{Comenius University, 842 48 Bratislava, Slovakia; Institute of Experimental Physics, 040 01 Kosice, Slovakia}
\author{G.~Apollinari$^{\dag}$}~\affiliation{Fermi National Accelerator Laboratory, Batavia, Illinois 60510, USA}
\author{J.A.~Appel$^{\dag}$}~\affiliation{Fermi National Accelerator Laboratory, Batavia, Illinois 60510, USA}
\author{T.~Arisawa$^{\dag}$}~\affiliation{Waseda University, Tokyo 169, Japan}
\author{A.~Artikov$^{\dag}$}~\affiliation{Joint Institute for Nuclear Research, Dubna, Russia}
\author{J.~Asaadi$^{\dag}$}~\affiliation{Texas A\&M University, College Station, Texas 77843, USA}
\author{W.~Ashmanskas$^{\dag}$}~\affiliation{Fermi National Accelerator Laboratory, Batavia, Illinois 60510, USA}
\author{A.~Askew$^{\ddag}$}~\affiliation{Florida State University, Tallahassee, Florida 32306, USA}
\author{S.~Atkins$^{\ddag}$}~\affiliation{Louisiana Tech University, Ruston, Louisiana 71272, USA}
\author{B.~Auerbach$^{\dag}$}~\affiliation{Yale University, New Haven, Connecticut 06520, USA}
\author{K.~Augsten$^{\ddag}$}~\affiliation{Czech Technical University in Prague, Prague, Czech Republic}
\author{A.~Aurisano$^{\dag}$}~\affiliation{Texas A\&M University, College Station, Texas 77843, USA}
\author{C.~Avila$^{\ddag}$}~\affiliation{Universidad de los Andes, Bogot\'a, Colombia}
\author{F.~Azfar$^{\dag}$}~\affiliation{University of Oxford, Oxford OX1 3RH, United Kingdom}
\author{F.~Badaud$^{\ddag}$}~\affiliation{LPC, Universit\'e Blaise Pascal, CNRS/IN2P3, Clermont, France}
\author{W.~Badgett$^{\dag}$}~\affiliation{Fermi National Accelerator Laboratory, Batavia, Illinois 60510, USA}
\author{T.~Bae$^{\dag}$}~\affiliation{Center for High Energy Physics: Kyungpook National University, Daegu 702-701, Korea; Seoul National University, Seoul 151-742, Korea; Sungkyunkwan University, Suwon 440-746, Korea; Korea Institute of Science and Technology Information, Daejeon 305-806, Korea; Chonnam National University, Gwangju 500-757, Korea; Chonbuk National University, Jeonju 561-756, Korea}
\author{L.~Bagby$^{\ddag}$}~\affiliation{Fermi National Accelerator Laboratory, Batavia, Illinois 60510, USA}
\author{B.~Baldin$^{\ddag}$}~\affiliation{Fermi National Accelerator Laboratory, Batavia, Illinois 60510, USA}
\author{D.V.~Bandurin$^{\ddag}$}~\affiliation{Florida State University, Tallahassee, Florida 32306, USA}
\author{S.~Banerjee$^{\ddag}$}~\affiliation{Tata Institute of Fundamental Research, Mumbai, India}
\author{A.~Barbaro-Galtieri$^{\dag}$}~\affiliation{Ernest Orlando Lawrence Berkeley National Laboratory, Berkeley, California 94720, USA}
\author{E.~Barberis$^{\ddag}$}~\affiliation{Northeastern University, Boston, Massachusetts 02115, USA}
\author{P.~Baringer$^{\ddag}$}~\affiliation{University of Kansas, Lawrence, Kansas 66045, USA}
\author{V.E.~Barnes$^{\dag}$}~\affiliation{Purdue University, West Lafayette, Indiana 47907, USA}
\author{B.A.~Barnett$^{\dag}$}~\affiliation{The Johns Hopkins University, Baltimore, Maryland 21218, USA}
\author{P.~Barria$^{\dag\sharp{d}}$}~\affiliation{Istituto Nazionale di Fisica Nucleare Pisa, $^{\sharp{c}}$University of Pisa, $^{\sharp{d}}$University of Siena and $^{\sharp{e}}$Scuola Normale Superiore, I-56127 Pisa, Italy}
\author{J.F.~Bartlett$^{\ddag}$}~\affiliation{Fermi National Accelerator Laboratory, Batavia, Illinois 60510, USA}
\author{P.~Bartos$^{\dag}$}~\affiliation{Comenius University, 842 48 Bratislava, Slovakia; Institute of Experimental Physics, 040 01 Kosice, Slovakia}
\author{U.~Bassler$^{\ddag}$}~\affiliation{CEA, Irfu, SPP, Saclay, France}
\author{M.~Bauce$^{\dag\sharp{b}}$}~\affiliation{Istituto Nazionale di Fisica Nucleare, Sezione di Padova-Trento, $^{\sharp{b}}$University of Padova, I-35131 Padova, Italy}
\author{V.~Bazterra$^{\ddag}$}~\affiliation{University of Illinois at Chicago, Chicago, Illinois 60607, USA}
\author{A.~Bean$^{\ddag}$}~\affiliation{University of Kansas, Lawrence, Kansas 66045, USA}
\author{F.~Bedeschi$^{\dag}$}~\affiliation{Istituto Nazionale di Fisica Nucleare Pisa, $^{\sharp{c}}$University of Pisa, $^{\sharp{d}}$University of Siena and $^{\sharp{e}}$Scuola Normale Superiore, I-56127 Pisa, Italy}
\author{M.~Begalli$^{\ddag}$}~\affiliation{Universidade do Estado do Rio de Janeiro, Rio de Janeiro, Brazil}
\author{S.~Behari$^{\dag}$}~\affiliation{The Johns Hopkins University, Baltimore, Maryland 21218, USA}
\author{L.~Bellantoni$^{\ddag}$}~\affiliation{Fermi National Accelerator Laboratory, Batavia, Illinois 60510, USA}
\author{G.~Bellettini$^{\dag\sharp{c}}$}~\affiliation{Istituto Nazionale di Fisica Nucleare Pisa, $^{\sharp{c}}$University of Pisa, $^{\sharp{d}}$University of Siena and $^{\sharp{e}}$Scuola Normale Superiore, I-56127 Pisa, Italy}
\author{J.~Bellinger$^{\dag}$}~\affiliation{University of Wisconsin, Madison, Wisconsin 53706, USA}
\author{D.~Benjamin$^{\dag}$}~\affiliation{Duke University, Durham, North Carolina 27708, USA}
\author{A.~Beretvas$^{\dag}$}~\affiliation{Fermi National Accelerator Laboratory, Batavia, Illinois 60510, USA}
\author{S.B.~Beri$^{\ddag}$}~\affiliation{Panjab University, Chandigarh, India}
\author{G.~Bernardi$^{\ddag}$}~\affiliation{LPNHE, Universit\'es Paris VI and VII, CNRS/IN2P3, Paris, France}
\author{R.~Bernhard$^{\ddag}$}~\affiliation{Physikalisches Institut, Universit\"at Freiburg, Freiburg, Germany}
\author{I.~Bertram$^{\ddag}$}~\affiliation{Lancaster University, Lancaster LA1 4YB, United Kingdom}
\author{M.~Besan\c{c}on$^{\ddag}$}~\affiliation{CEA, Irfu, SPP, Saclay, France}
\author{R.~Beuselinck$^{\ddag}$}~\affiliation{Imperial College London, London SW7 2AZ, United Kingdom}
\author{P.C.~Bhat$^{\ddag}$}~\affiliation{Fermi National Accelerator Laboratory, Batavia, Illinois 60510, USA}
\author{S.~Bhatia$^{\ddag}$}~\affiliation{University of Mississippi, University, Mississippi 38677, USA}
\author{V.~Bhatnagar$^{\ddag}$}~\affiliation{Panjab University, Chandigarh, India}
\author{A.~Bhatti$^{\dag}$}~\affiliation{The Rockefeller University, New York, New York 10065, USA}
\author{D.~Bisello$^{\dag\sharp{b}}$}~\affiliation{Istituto Nazionale di Fisica Nucleare, Sezione di Padova-Trento, $^{\sharp{b}}$University of Padova, I-35131 Padova, Italy}
\author{I.~Bizjak$^{\dag}$}~\affiliation{University College London, London WC1E 6BT, United Kingdom}
\author{K.R.~Bland$^{\dag}$}~\affiliation{Baylor University, Waco, Texas 76798, USA}
\author{G.~Blazey$^{\ddag}$}~\affiliation{Northern Illinois University, DeKalb, Illinois 60115, USA}
\author{S.~Blessing$^{\ddag}$}~\affiliation{Florida State University, Tallahassee, Florida 32306, USA}
\author{K.~Bloom$^{\ddag}$}~\affiliation{University of Nebraska, Lincoln, Nebraska 68588, USA}
\author{B.~Blumenfeld$^{\dag}$}~\affiliation{The Johns Hopkins University, Baltimore, Maryland 21218, USA}
\author{A.~Bocci$^{\dag}$}~\affiliation{Duke University, Durham, North Carolina 27708, USA}
\author{A.~Bodek$^{\dag}$}~\affiliation{University of Rochester, Rochester, New York 14627, USA}
\author{A.~Boehnlein$^{\ddag}$}~\affiliation{Fermi National Accelerator Laboratory, Batavia, Illinois 60510, USA}
\author{D.~Boline$^{\ddag}$}~\affiliation{State University of New York, Stony Brook, New York 11794, USA}
\author{E.E.~Boos$^{\ddag}$}~\affiliation{Moscow State University, Moscow, Russia}
\author{G.~Borissov$^{\ddag}$}~\affiliation{Lancaster University, Lancaster LA1 4YB, United Kingdom}
\author{D.~Bortoletto$^{\dag}$}~\affiliation{Purdue University, West Lafayette, Indiana 47907, USA}
\author{T.~Bose$^{\ddag}$}~\affiliation{Boston University, Boston, Massachusetts 02215, USA}
\author{J.~Boudreau$^{\dag}$}~\affiliation{University of Pittsburgh, Pittsburgh, Pennsylvania 15260, USA}
\author{A.~Boveia$^{\dag}$}~\affiliation{Enrico Fermi Institute, University of Chicago, Chicago, Illinois 60637, USA}
\author{A.~Brandt$^{\ddag}$}~\affiliation{University of Texas, Arlington, Texas 76019, USA}
\author{O.~Brandt$^{\ddag}$}~\affiliation{II. Physikalisches Institut, Georg-August-Universit\"at G\"ottingen, G\"ottingen, Germany}
\author{L.~Brigliadori$^{\dag\sharp{a}}$}~\affiliation{Istituto Nazionale di Fisica Nucleare Bologna, $^{\sharp{a}}$University of Bologna, I-40127 Bologna, Italy}
\author{R.~Brock$^{\ddag}$}~\affiliation{Michigan State University, East Lansing, Michigan 48824, USA}
\author{C.~Bromberg$^{\dag}$}~\affiliation{Michigan State University, East Lansing, Michigan 48824, USA}
\author{A.~Bross$^{\ddag}$}~\affiliation{Fermi National Accelerator Laboratory, Batavia, Illinois 60510, USA}
\author{D.~Brown$^{\ddag}$}~\affiliation{LPNHE, Universit\'es Paris VI and VII, CNRS/IN2P3, Paris, France}
\author{J.~Brown$^{\ddag}$}~\affiliation{LPNHE, Universit\'es Paris VI and VII, CNRS/IN2P3, Paris, France}
\author{E.~Brucken$^{\dag}$}~\affiliation{Division of High Energy Physics, Department of Physics, University of Helsinki and Helsinki Institute of Physics, FIN-00014, Helsinki, Finland}
\author{J.~Budagov$^{\dag}$}~\affiliation{Joint Institute for Nuclear Research, Dubna, Russia}
\author{X.B.~Bu$^{\ddag}$}~\affiliation{Fermi National Accelerator Laboratory, Batavia, Illinois 60510, USA}
\author{H.S.~Budd$^{\dag}$}~\affiliation{University of Rochester, Rochester, New York 14627, USA}
\author{M.~Buehler$^{\ddag}$}~\affiliation{Fermi National Accelerator Laboratory, Batavia, Illinois 60510, USA}
\author{V.~Buescher$^{\ddag}$}~\affiliation{Institut f\"ur Physik, Universit\"at Mainz, Mainz, Germany}
\author{V.~Bunichev$^{\ddag}$}~\affiliation{Moscow State University, Moscow, Russia}
\author{S.~Burdin$^{\ddag b}$}~\affiliation{Lancaster University, Lancaster LA1 4YB, United Kingdom}
\author{K.~Burkett$^{\dag}$}~\affiliation{Fermi National Accelerator Laboratory, Batavia, Illinois 60510, USA}
\author{G.~Busetto$^{\dag\sharp{b}}$}~\affiliation{Istituto Nazionale di Fisica Nucleare, Sezione di Padova-Trento, $^{\sharp{b}}$University of Padova, I-35131 Padova, Italy}
\author{P.~Bussey$^{\dag}$}~\affiliation{Glasgow University, Glasgow G12 8QQ, United Kingdom}
\author{C.P.~Buszello$^{\ddag}$}~\affiliation{Uppsala University, Uppsala, Sweden}
\author{A.~Buzatu$^{\dag}$}~\affiliation{Institute of Particle Physics: McGill University, Montr\'{e}al, Qu\'{e}bec, Canada H3A~2T8; Simon Fraser University, Burnaby, British Columbia, Canada V5A~1S6; University of Toronto, Toronto, Ontario, Canada M5S~1A7; and TRIUMF, Vancouver, British Columbia, Canada V6T~2A3}
\author{A.~Calamba$^{\dag}$}~\affiliation{Carnegie Mellon University, Pittsburgh, Pennsylvania 15213, USA}
\author{C.~Calancha$^{\dag}$}~\affiliation{Centro de Investigaciones Energeticas Medioambientales y Tecnologicas, E-28040 Madrid, Spain}
\author{E.~Camacho-P\'erez$^{\ddag}$}~\affiliation{CINVESTAV, Mexico City, Mexico}
\author{S.~Camarda$^{\dag}$}~\affiliation{Institut de Fisica d'Altes Energies, ICREA, Universitat Autonoma de Barcelona, E-08193, Bellaterra (Barcelona), Spain}
\author{M.~Campanelli$^{\dag}$}~\affiliation{University College London, London WC1E 6BT, United Kingdom}
\author{M.~Campbell$^{\dag}$}~\affiliation{University of Michigan, Ann Arbor, Michigan 48109, USA}
\author{F.~Canelli$^{\dag}$}~\affiliation{Enrico Fermi Institute, University of Chicago, Chicago, Illinois 60637, USA}
\author{B.~Carls$^{\dag}$}~\affiliation{University of Illinois, Urbana, Illinois 61801, USA}
\author{D.~Carlsmith$^{\dag}$}~\affiliation{University of Wisconsin, Madison, Wisconsin 53706, USA}
\author{R.~Carosi$^{\dag}$}~\affiliation{Istituto Nazionale di Fisica Nucleare Pisa, $^{\sharp{c}}$University of Pisa, $^{\sharp{d}}$University of Siena and $^{\sharp{e}}$Scuola Normale Superiore, I-56127 Pisa, Italy}
\author{S.~Carrillo$^{\dag c}$}~\affiliation{University of Florida, Gainesville, Florida 32611, USA}
\author{S.~Carron$^{\dag}$}~\affiliation{Fermi National Accelerator Laboratory, Batavia, Illinois 60510, USA}
\author{B.~Casal$^{\dag d}$}~\affiliation{Instituto de Fisica de Cantabria, CSIC-University of Cantabria, 39005 Santander, Spain}
\author{M.~Casarsa$^{\dag}$}~\affiliation{Istituto Nazionale di Fisica Nucleare Trieste/Udine, I-34100 Trieste, $^{\sharp{g}}$University of Udine, I-33100 Udine, Italy}
\author{B.C.K.~Casey$^{\ddag}$}~\affiliation{Fermi National Accelerator Laboratory, Batavia, Illinois 60510, USA}
\author{H.~Castilla-Valdez$^{\ddag}$}~\affiliation{CINVESTAV, Mexico City, Mexico}
\author{A.~Castro$^{\dag\sharp{a}}$}~\affiliation{Istituto Nazionale di Fisica Nucleare Bologna, $^{\sharp{a}}$University of Bologna, I-40127 Bologna, Italy}
\author{P.~Catastini$^{\dag}$}~\affiliation{Harvard University, Cambridge, Massachusetts 02138, USA}
\author{S.~Caughron$^{\ddag}$}~\affiliation{Michigan State University, East Lansing, Michigan 48824, USA}
\author{D.~Cauz$^{\dag}$}~\affiliation{Istituto Nazionale di Fisica Nucleare Trieste/Udine, I-34100 Trieste, $^{\sharp{g}}$University of Udine, I-33100 Udine, Italy}
\author{V.~Cavaliere$^{\dag}$}~\affiliation{University of Illinois, Urbana, Illinois 61801, USA}
\author{M.~Cavalli-Sforza$^{\dag}$}~\affiliation{Institut de Fisica d'Altes Energies, ICREA, Universitat Autonoma de Barcelona, E-08193, Bellaterra (Barcelona), Spain}
\author{A.~Cerri$^{\dag e}$}~\affiliation{Ernest Orlando Lawrence Berkeley National Laboratory, Berkeley, California 94720, USA}
\author{L.~Cerrito$^{\dag f}$}~\affiliation{University College London, London WC1E 6BT, United Kingdom}
\author{S.~Chakrabarti$^{\ddag}$}~\affiliation{State University of New York, Stony Brook, New York 11794, USA}
\author{D.~Chakraborty$^{\ddag}$}~\affiliation{Northern Illinois University, DeKalb, Illinois 60115, USA}
\author{K.M.~Chan$^{\ddag}$}~\affiliation{University of Notre Dame, Notre Dame, Indiana 46556, USA}
\author{A.~Chandra$^{\ddag}$}~\affiliation{Rice University, Houston, Texas 77005, USA}
\author{E.~Chapon$^{\ddag}$}~\affiliation{CEA, Irfu, SPP, Saclay, France}
\author{G.~Chen$^{\ddag}$}~\affiliation{University of Kansas, Lawrence, Kansas 66045, USA}
\author{Y.C.~Chen$^{\dag}$}~\affiliation{Institute of Physics, Academia Sinica, Taipei, Taiwan 11529, Republic of China}
\author{M.~Chertok$^{\dag}$}~\affiliation{University of California, Davis, Davis, California 95616, USA}
\author{S.~Chevalier-Th\'ery$^{\ddag}$}~\affiliation{CEA, Irfu, SPP, Saclay, France}
\author{G.~Chiarelli$^{\dag}$}~\affiliation{Istituto Nazionale di Fisica Nucleare Pisa, $^{\sharp{c}}$University of Pisa, $^{\sharp{d}}$University of Siena and $^{\sharp{e}}$Scuola Normale Superiore, I-56127 Pisa, Italy}
\author{G.~Chlachidze$^{\dag}$}~\affiliation{Fermi National Accelerator Laboratory, Batavia, Illinois 60510, USA}
\author{F.~Chlebana$^{\dag}$}~\affiliation{Fermi National Accelerator Laboratory, Batavia, Illinois 60510, USA}
\author{D.K.~Cho$^{\ddag}$}~\affiliation{Brown University, Providence, Rhode Island 02912, USA}
\author{K.~Cho$^{\dag}$}~\affiliation{Center for High Energy Physics: Kyungpook National University, Daegu 702-701, Korea; Seoul National University, Seoul 151-742, Korea; Sungkyunkwan University, Suwon 440-746, Korea; Korea Institute of Science and Technology Information, Daejeon 305-806, Korea; Chonnam National University, Gwangju 500-757, Korea; Chonbuk National University, Jeonju 561-756, Korea}
\author{S.W.~Cho$^{\ddag}$}~\affiliation{Korea Detector Laboratory, Korea University, Seoul, Korea}
\author{S.~Choi$^{\ddag}$}~\affiliation{Korea Detector Laboratory, Korea University, Seoul, Korea}
\author{D.~Chokheli$^{\dag}$}~\affiliation{Joint Institute for Nuclear Research, Dubna, Russia}
\author{B.~Choudhary$^{\ddag}$}~\affiliation{Delhi University, Delhi, India}
\author{W.H.~Chung$^{\dag}$}~\affiliation{University of Wisconsin, Madison, Wisconsin 53706, USA}
\author{Y.S.~Chung$^{\dag}$}~\affiliation{University of Rochester, Rochester, New York 14627, USA}
\author{S.~Cihangir$^{\ddag}$}~\affiliation{Fermi National Accelerator Laboratory, Batavia, Illinois 60510, USA}
\author{M.A.~Ciocci$^{\dag\sharp{d}}$}~\affiliation{Istituto Nazionale di Fisica Nucleare Pisa, $^{\sharp{c}}$University of Pisa, $^{\sharp{d}}$University of Siena and $^{\sharp{e}}$Scuola Normale Superiore, I-56127 Pisa, Italy}
\author{D.~Claes$^{\ddag}$}~\affiliation{University of Nebraska, Lincoln, Nebraska 68588, USA}
\author{A.~Clark$^{\dag}$}~\affiliation{University of Geneva, CH-1211 Geneva 4, Switzerland}
\author{C.~Clarke$^{\dag}$}~\affiliation{Wayne State University, Detroit, Michigan 48201, USA}
\author{J.~Clutter$^{\ddag}$}~\affiliation{University of Kansas, Lawrence, Kansas 66045, USA}
\author{G.~Compostella$^{\dag\sharp{b}}$}~\affiliation{Istituto Nazionale di Fisica Nucleare, Sezione di Padova-Trento, $^{\sharp{b}}$University of Padova, I-35131 Padova, Italy}
\author{M.E.~Convery$^{\dag}$}~\affiliation{Fermi National Accelerator Laboratory, Batavia, Illinois 60510, USA}
\author{J.~Conway$^{\dag}$}~\affiliation{University of California, Davis, Davis, California 95616, USA}
\author{M.~Cooke$^{\ddag}$}~\affiliation{Fermi National Accelerator Laboratory, Batavia, Illinois 60510, USA}
\author{W.E.~Cooper$^{\ddag}$}~\affiliation{Fermi National Accelerator Laboratory, Batavia, Illinois 60510, USA}
\author{M.~Corbo$^{\dag}$}~\affiliation{Fermi National Accelerator Laboratory, Batavia, Illinois 60510, USA}
\author{M.~Corcoran$^{\ddag}$}~\affiliation{Rice University, Houston, Texas 77005, USA}
\author{M.~Cordelli$^{\dag}$}~\affiliation{Laboratori Nazionali di Frascati, Istituto Nazionale di Fisica Nucleare, I-00044 Frascati, Italy}
\author{F.~Couderc$^{\ddag}$}~\affiliation{CEA, Irfu, SPP, Saclay, France}
\author{M.-C.~Cousinou$^{\ddag}$}~\affiliation{CPPM, Aix-Marseille Universit\'e, CNRS/IN2P3, Marseille, France}
\author{C.A.~Cox$^{\dag}$}~\affiliation{University of California, Davis, Davis, California 95616, USA}
\author{D.J.~Cox$^{\dag}$}~\affiliation{University of California, Davis, Davis, California 95616, USA}
\author{F.~Crescioli$^{\dag\sharp{c}}$}~\affiliation{Istituto Nazionale di Fisica Nucleare Pisa, $^{\sharp{c}}$University of Pisa, $^{\sharp{d}}$University of Siena and $^{\sharp{e}}$Scuola Normale Superiore, I-56127 Pisa, Italy}
\author{A.~Croc$^{\ddag}$}~\affiliation{CEA, Irfu, SPP, Saclay, France}
\author{J.~Cuevas$^{\dag a}$}~\affiliation{Instituto de Fisica de Cantabria, CSIC-University of Cantabria, 39005 Santander, Spain}
\author{R.~Culbertson$^{\dag}$}~\affiliation{Fermi National Accelerator Laboratory, Batavia, Illinois 60510, USA}
\author{D.~Cutts$^{\ddag}$}~\affiliation{Brown University, Providence, Rhode Island 02912, USA}
\author{D.~Dagenhart$^{\dag}$}~\affiliation{Fermi National Accelerator Laboratory, Batavia, Illinois 60510, USA}
\author{N.~d'Ascenzo$^{\dag g}$}~\affiliation{Fermi National Accelerator Laboratory, Batavia, Illinois 60510, USA}
\author{A.~Das$^{\ddag}$}~\affiliation{University of Arizona, Tucson, Arizona 85721, USA}
\author{M.~Datta$^{\dag}$}~\affiliation{Fermi National Accelerator Laboratory, Batavia, Illinois 60510, USA}
\author{G.~Davies$^{\ddag}$}~\affiliation{Imperial College London, London SW7 2AZ, United Kingdom}
\author{P.~de~Barbaro$^{\dag}$}~\affiliation{University of Rochester, Rochester, New York 14627, USA}
\author{S.J.~de~Jong$^{\ddag}$}~\affiliation{Nikhef, Science Park, Amsterdam, the Netherlands}~\affiliation{Radboud University Nijmegen, Nijmegen, the Netherlands}
\author{E.~De~La~Cruz-Burelo$^{\ddag}$}~\affiliation{CINVESTAV, Mexico City, Mexico}
\author{F.~D\'eliot$^{\ddag}$}~\affiliation{CEA, Irfu, SPP, Saclay, France}
\author{M.~Dell'Orso$^{\dag\sharp{c}}$}~\affiliation{Istituto Nazionale di Fisica Nucleare Pisa, $^{\sharp{c}}$University of Pisa, $^{\sharp{d}}$University of Siena and $^{\sharp{e}}$Scuola Normale Superiore, I-56127 Pisa, Italy}
\author{R.~Demina$^{\ddag}$}~\affiliation{University of Rochester, Rochester, New York 14627, USA}
\author{L.~Demortier$^{\dag}$}~\affiliation{The Rockefeller University, New York, New York 10065, USA}
\author{M.~Deninno$^{\dag}$}~\affiliation{Istituto Nazionale di Fisica Nucleare Bologna, $^{\sharp{a}}$University of Bologna, I-40127 Bologna, Italy}
\author{D.~Denisov$^{\ddag}$}~\affiliation{Fermi National Accelerator Laboratory, Batavia, Illinois 60510, USA}
\author{S.P.~Denisov$^{\ddag}$}~\affiliation{Institute for High Energy Physics, Protvino, Russia}
\author{M.~d'Errico$^{\dag\sharp{b}}$}~\affiliation{Istituto Nazionale di Fisica Nucleare, Sezione di Padova-Trento, $^{\sharp{b}}$University of Padova, I-35131 Padova, Italy}
\author{S.~Desai$^{\ddag}$}~\affiliation{Fermi National Accelerator Laboratory, Batavia, Illinois 60510, USA}
\author{C.~Deterre$^{\ddag}$}~\affiliation{CEA, Irfu, SPP, Saclay, France}
\author{K.~DeVaughan$^{\ddag}$}~\affiliation{University of Nebraska, Lincoln, Nebraska 68588, USA}
\author{F.~Devoto$^{\dag}$}~\affiliation{Division of High Energy Physics, Department of Physics, University of Helsinki and Helsinki Institute of Physics, FIN-00014, Helsinki, Finland}
\author{A.~Di~Canto$^{\dag\sharp{c}}$}~\affiliation{Istituto Nazionale di Fisica Nucleare Pisa, $^{\sharp{c}}$University of Pisa, $^{\sharp{d}}$University of Siena and $^{\sharp{e}}$Scuola Normale Superiore, I-56127 Pisa, Italy}
\author{B.~Di~Ruzza$^{\dag}$}~\affiliation{Fermi National Accelerator Laboratory, Batavia, Illinois 60510, USA}
\author{H.T.~Diehl$^{\ddag}$}~\affiliation{Fermi National Accelerator Laboratory, Batavia, Illinois 60510, USA}
\author{M.~Diesburg$^{\ddag}$}~\affiliation{Fermi National Accelerator Laboratory, Batavia, Illinois 60510, USA}
\author{P.F.~Ding$^{\ddag}$}~\affiliation{The University of Manchester, Manchester M13 9PL, United Kingdom}
\author{J.R.~Dittmann$^{\dag}$}~\affiliation{Baylor University, Waco, Texas 76798, USA}
\author{A.~Dominguez$^{\ddag}$}~\affiliation{University of Nebraska, Lincoln, Nebraska 68588, USA}
\author{S.~Donati$^{\dag\sharp{c}}$}~\affiliation{Istituto Nazionale di Fisica Nucleare Pisa, $^{\sharp{c}}$University of Pisa, $^{\sharp{d}}$University of Siena and $^{\sharp{e}}$Scuola Normale Superiore, I-56127 Pisa, Italy}
\author{P.~Dong$^{\dag}$}~\affiliation{Fermi National Accelerator Laboratory, Batavia, Illinois 60510, USA}
\author{M.~D'Onofrio$^{\dag}$}~\affiliation{University of Liverpool, Liverpool L69 7ZE, United Kingdom}
\author{M.~Dorigo$^{\dag}$}~\affiliation{Istituto Nazionale di Fisica Nucleare Trieste/Udine, I-34100 Trieste, $^{\sharp{g}}$University of Udine, I-33100 Udine, Italy}
\author{T.~Dorigo$^{\dag}$}~\affiliation{Istituto Nazionale di Fisica Nucleare, Sezione di Padova-Trento, $^{\sharp{b}}$University of Padova, I-35131 Padova, Italy}
\author{A.~Dubey$^{\ddag}$}~\affiliation{Delhi University, Delhi, India}
\author{L.V.~Dudko$^{\ddag}$}~\affiliation{Moscow State University, Moscow, Russia}
\author{D.~Duggan$^{\ddag}$}~\affiliation{Rutgers University, Piscataway, New Jersey 08855, USA}
\author{A.~Duperrin$^{\ddag}$}~\affiliation{CPPM, Aix-Marseille Universit\'e, CNRS/IN2P3, Marseille, France}
\author{S.~Dutt$^{\ddag}$}~\affiliation{Panjab University, Chandigarh, India}
\author{A.~Dyshkant$^{\ddag}$}~\affiliation{Northern Illinois University, DeKalb, Illinois 60115, USA}
\author{M.~Eads$^{\ddag}$}~\affiliation{University of Nebraska, Lincoln, Nebraska 68588, USA}
\author{K.~Ebina$^{\dag}$}~\affiliation{Waseda University, Tokyo 169, Japan}
\author{D.~Edmunds$^{\ddag}$}~\affiliation{Michigan State University, East Lansing, Michigan 48824, USA}
\author{A.~Elagin$^{\dag}$}~\affiliation{Texas A\&M University, College Station, Texas 77843, USA}
\author{J.~Ellison$^{\ddag}$}~\affiliation{University of California Riverside, Riverside, California 92521, USA}
\author{V.D.~Elvira$^{\ddag}$}~\affiliation{Fermi National Accelerator Laboratory, Batavia, Illinois 60510, USA}
\author{Y.~Enari$^{\ddag}$}~\affiliation{LPNHE, Universit\'es Paris VI and VII, CNRS/IN2P3, Paris, France}
\author{A.~Eppig$^{\dag}$}~\affiliation{University of Michigan, Ann Arbor, Michigan 48109, USA}
\author{R.~Erbacher$^{\dag}$}~\affiliation{University of California, Davis, Davis, California 95616, USA}
\author{S.~Errede$^{\dag}$}~\affiliation{University of Illinois, Urbana, Illinois 61801, USA}
\author{N.~Ershaidat$^{\dag h}$}~\affiliation{Fermi National Accelerator Laboratory, Batavia, Illinois 60510, USA}
\author{R.~Eusebi$^{\dag}$}~\affiliation{Texas A\&M University, College Station, Texas 77843, USA}
\author{H.~Evans$^{\ddag}$}~\affiliation{Indiana University, Bloomington, Indiana 47405, USA}
\author{A.~Evdokimov$^{\ddag}$}~\affiliation{Brookhaven National Laboratory, Upton, New York 11973, USA}
\author{V.N.~Evdokimov$^{\ddag}$}~\affiliation{Institute for High Energy Physics, Protvino, Russia}
\author{G.~Facini$^{\ddag}$}~\affiliation{Northeastern University, Boston, Massachusetts 02115, USA}
\author{S.~Farrington$^{\dag}$}~\affiliation{University of Oxford, Oxford OX1 3RH, United Kingdom}
\author{M.~Feindt$^{\dag}$}~\affiliation{Institut f\"{u}r Experimentelle Kernphysik, Karlsruhe Institute of Technology, D-76131 Karlsruhe, Germany}
\author{L.~Feng$^{\ddag}$}~\affiliation{Northern Illinois University, DeKalb, Illinois 60115, USA}
\author{T.~Ferbel$^{\ddag}$}~\affiliation{University of Rochester, Rochester, New York 14627, USA}
\author{J.P.~Fernandez$^{\dag}$}~\affiliation{Centro de Investigaciones Energeticas Medioambientales y Tecnologicas, E-28040 Madrid, Spain}
\author{F.~Fiedler$^{\ddag}$}~\affiliation{Institut f\"ur Physik, Universit\"at Mainz, Mainz, Germany}
\author{R.~Field$^{\dag}$}~\affiliation{University of Florida, Gainesville, Florida 32611, USA}
\author{F.~Filthaut$^{\ddag}$}~\affiliation{Nikhef, Science Park, Amsterdam, the Netherlands}~\affiliation{Radboud University Nijmegen, Nijmegen, the Netherlands}
\author{W.~Fisher$^{\ddag}$}~\affiliation{Michigan State University, East Lansing, Michigan 48824, USA}
\author{H.E.~Fisk$^{\ddag}$}~\affiliation{Fermi National Accelerator Laboratory, Batavia, Illinois 60510, USA}
\author{G.~Flanagan$^{\dag i}$}~\affiliation{Fermi National Accelerator Laboratory, Batavia, Illinois 60510, USA}
\author{R.~Forrest$^{\dag}$}~\affiliation{University of California, Davis, Davis, California 95616, USA}
\author{M.~Fortner$^{\ddag}$}~\affiliation{Northern Illinois University, DeKalb, Illinois 60115, USA}
\author{H.~Fox$^{\ddag}$}~\affiliation{Lancaster University, Lancaster LA1 4YB, United Kingdom}
\author{M.J.~Frank$^{\dag}$}~\affiliation{Baylor University, Waco, Texas 76798, USA}
\author{M.~Franklin$^{\dag}$}~\affiliation{Harvard University, Cambridge, Massachusetts 02138, USA}
\author{J.C.~Freeman$^{\dag}$}~\affiliation{Fermi National Accelerator Laboratory, Batavia, Illinois 60510, USA}
\author{S.~Fuess$^{\ddag}$}~\affiliation{Fermi National Accelerator Laboratory, Batavia, Illinois 60510, USA}
\author{Y.~Funakoshi$^{\dag}$}~\affiliation{Waseda University, Tokyo 169, Japan}
\author{I.~Furic$^{\dag}$}~\affiliation{University of Florida, Gainesville, Florida 32611, USA}
\author{M.~Gallinaro$^{\dag}$}~\affiliation{The Rockefeller University, New York, New York 10065, USA}
\author{A.~Garcia-Bellido$^{\ddag}$}~\affiliation{University of Rochester, Rochester, New York 14627, USA}
\author{J.E.~Garcia$^{\dag}$}~\affiliation{University of Geneva, CH-1211 Geneva 4, Switzerland}
\author{J.A.~Garc\'{\i}a-Gonz\'alez$^{\ddag}$}~\affiliation{CINVESTAV, Mexico City, Mexico}
\author{G.A.~Garc\'ia-Guerra$^{\ddag c}$}~\affiliation{CINVESTAV, Mexico City, Mexico}
\author{A.F.~Garfinkel$^{\dag}$}~\affiliation{Purdue University, West Lafayette, Indiana 47907, USA}
\author{P.~Garosi$^{\dag\sharp{d}}$}~\affiliation{Istituto Nazionale di Fisica Nucleare Pisa, $^{\sharp{c}}$University of Pisa, $^{\sharp{d}}$University of Siena and $^{\sharp{e}}$Scuola Normale Superiore, I-56127 Pisa, Italy}
\author{V.~Gavrilov$^{\ddag}$}~\affiliation{Institute for Theoretical and Experimental Physics, Moscow, Russia}
\author{P.~Gay$^{\ddag}$}~\affiliation{LPC, Universit\'e Blaise Pascal, CNRS/IN2P3, Clermont, France}
\author{W.~Geng$^{\ddag}$}~\affiliation{CPPM, Aix-Marseille Universit\'e, CNRS/IN2P3, Marseille, France}~\affiliation{Michigan State University, East Lansing, Michigan 48824, USA}
\author{D.~Gerbaudo$^{\ddag}$}~\affiliation{Princeton University, Princeton, New Jersey 08544, USA}
\author{C.E.~Gerber$^{\ddag}$}~\affiliation{University of Illinois at Chicago, Chicago, Illinois 60607, USA}
\author{H.~Gerberich$^{\dag}$}~\affiliation{University of Illinois, Urbana, Illinois 61801, USA}
\author{E.~Gerchtein$^{\dag}$}~\affiliation{Fermi National Accelerator Laboratory, Batavia, Illinois 60510, USA}
\author{Y.~Gershtein$^{\ddag}$}~\affiliation{Rutgers University, Piscataway, New Jersey 08855, USA}
\author{S.~Giagu$^{\dag}$}~\affiliation{Istituto Nazionale di Fisica Nucleare, Sezione di Roma 1, $^{\sharp{f}}$Sapienza Universit\`{a} di Roma, I-00185 Roma, Italy}
\author{V.~Giakoumopoulou$^{\dag}$}~\affiliation{University of Athens, 157 71 Athens, Greece}
\author{P.~Giannetti$^{\dag}$}~\affiliation{Istituto Nazionale di Fisica Nucleare Pisa, $^{\sharp{c}}$University of Pisa, $^{\sharp{d}}$University of Siena and $^{\sharp{e}}$Scuola Normale Superiore, I-56127 Pisa, Italy}
\author{K.~Gibson$^{\dag}$}~\affiliation{University of Pittsburgh, Pittsburgh, Pennsylvania 15260, USA}
\author{C.M.~Ginsburg$^{\dag}$}~\affiliation{Fermi National Accelerator Laboratory, Batavia, Illinois 60510, USA}
\author{G.~Ginther$^{\ddag}$}~\affiliation{Fermi National Accelerator Laboratory, Batavia, Illinois 60510, USA}~\affiliation{University of Rochester, Rochester, New York 14627, USA}
\author{N.~Giokaris$^{\dag}$}~\affiliation{University of Athens, 157 71 Athens, Greece}
\author{P.~Giromini$^{\dag}$}~\affiliation{Laboratori Nazionali di Frascati, Istituto Nazionale di Fisica Nucleare, I-00044 Frascati, Italy}
\author{G.~Giurgiu$^{\dag}$}~\affiliation{The Johns Hopkins University, Baltimore, Maryland 21218, USA}
\author{V.~Glagolev$^{\dag}$}~\affiliation{Joint Institute for Nuclear Research, Dubna, Russia}
\author{D.~Glenzinski$^{\dag}$}~\affiliation{Fermi National Accelerator Laboratory, Batavia, Illinois 60510, USA}
\author{M.~Gold$^{\dag}$}~\affiliation{University of New Mexico, Albuquerque, New Mexico 87131, USA}
\author{D.~Goldin$^{\dag}$}~\affiliation{Texas A\&M University, College Station, Texas 77843, USA}
\author{N.~Goldschmidt$^{\dag}$}~\affiliation{University of Florida, Gainesville, Florida 32611, USA}
\author{A.~Golossanov$^{\dag}$}~\affiliation{Fermi National Accelerator Laboratory, Batavia, Illinois 60510, USA}
\author{G.~Golovanov$^{\ddag}$}~\affiliation{Joint Institute for Nuclear Research, Dubna, Russia}
\author{G.~Gomez-Ceballos$^{\dag}$}~\affiliation{Massachusetts Institute of Technology, Cambridge, Massachusetts 02139, USA}
\author{G.~Gomez$^{\dag}$}~\affiliation{Instituto de Fisica de Cantabria, CSIC-University of Cantabria, 39005 Santander, Spain}
\author{M.~Goncharov$^{\dag}$}~\affiliation{Massachusetts Institute of Technology, Cambridge, Massachusetts 02139, USA}
\author{O.~Gonz\'{a}lez$^{\dag}$}~\affiliation{Centro de Investigaciones Energeticas Medioambientales y Tecnologicas, E-28040 Madrid, Spain}
\author{I.~Gorelov$^{\dag}$}~\affiliation{University of New Mexico, Albuquerque, New Mexico 87131, USA}
\author{A.T.~Goshaw$^{\dag}$}~\affiliation{Duke University, Durham, North Carolina 27708, USA}
\author{K.~Goulianos$^{\dag}$}~\affiliation{The Rockefeller University, New York, New York 10065, USA}
\author{A.~Goussiou$^{\ddag}$}~\affiliation{University of Washington, Seattle, Washington 98195, USA}
\author{P.D.~Grannis$^{\ddag}$}~\affiliation{State University of New York, Stony Brook, New York 11794, USA}
\author{S.~Greder$^{\ddag}$}~\affiliation{IPHC, Universit\'e de Strasbourg, CNRS/IN2P3, Strasbourg, France}
\author{H.~Greenlee$^{\ddag}$}~\affiliation{Fermi National Accelerator Laboratory, Batavia, Illinois 60510, USA}
\author{G.~Grenier$^{\ddag}$}~\affiliation{IPNL, Universit\'e Lyon 1, CNRS/IN2P3, Villeurbanne, France and Universit\'e de Lyon, Lyon, France}
\author{S.~Grinstein$^{\dag}$}~\affiliation{Institut de Fisica d'Altes Energies, ICREA, Universitat Autonoma de Barcelona, E-08193, Bellaterra (Barcelona), Spain}
\author{Ph.~Gris$^{\ddag}$}~\affiliation{LPC, Universit\'e Blaise Pascal, CNRS/IN2P3, Clermont, France}
\author{J.-F.~Grivaz$^{\ddag}$}~\affiliation{LAL, Universit\'e Paris-Sud, CNRS/IN2P3, Orsay, France}
\author{A.~Grohsjean$^{\ddag d}$}~\affiliation{CEA, Irfu, SPP, Saclay, France}
\author{C.~Grosso-Pilcher$^{\dag}$}~\affiliation{Enrico Fermi Institute, University of Chicago, Chicago, Illinois 60637, USA}
\author{R.C.~Group$^{\dag}$}~\affiliation{University of Virginia, Charlottesville, Virginia 22904, USA}~\affiliation{Fermi National Accelerator Laboratory, Batavia, Illinois 60510, USA}
\author{S.~Gr\"unendahl$^{\ddag}$}~\affiliation{Fermi National Accelerator Laboratory, Batavia, Illinois 60510, USA}
\author{M.W.~Gr{\"u}newald$^{\ddag}$}~\affiliation{University College Dublin, Dublin, Ireland}
\author{T.~Guillemin$^{\ddag}$}~\affiliation{LAL, Universit\'e Paris-Sud, CNRS/IN2P3, Orsay, France}
\author{J.~Guimaraes~da~Costa$^{\dag}$}~\affiliation{Harvard University, Cambridge, Massachusetts 02138, USA}
\author{G.~Gutierrez$^{\ddag}$}~\affiliation{Fermi National Accelerator Laboratory, Batavia, Illinois 60510, USA}
\author{P.~Gutierrez$^{\ddag}$}~\affiliation{University of Oklahoma, Norman, Oklahoma 73019, USA}
\author{S.~Hagopian$^{\ddag}$}~\affiliation{Florida State University, Tallahassee, Florida 32306, USA}
\author{S.R.~Hahn$^{\dag}$}~\affiliation{Fermi National Accelerator Laboratory, Batavia, Illinois 60510, USA}
\author{J.~Haley$^{\ddag}$}~\affiliation{Northeastern University, Boston, Massachusetts 02115, USA}
\author{E.~Halkiadakis$^{\dag}$}~\affiliation{Rutgers University, Piscataway, New Jersey 08855, USA}
\author{A.~Hamaguchi$^{\dag}$}~\affiliation{Osaka City University, Osaka 588, Japan}
\author{J.Y.~Han$^{\dag}$}~\affiliation{University of Rochester, Rochester, New York 14627, USA}
\author{L.~Han$^{\ddag}$}~\affiliation{University of Science and Technology of China, Hefei, People's Republic of China}
\author{F.~Happacher$^{\dag}$}~\affiliation{Laboratori Nazionali di Frascati, Istituto Nazionale di Fisica Nucleare, I-00044 Frascati, Italy}
\author{K.~Hara$^{\dag}$}~\affiliation{University of Tsukuba, Tsukuba, Ibaraki 305, Japan}
\author{K.~Harder$^{\ddag}$}~\affiliation{The University of Manchester, Manchester M13 9PL, United Kingdom}
\author{D.~Hare$^{\dag}$}~\affiliation{Rutgers University, Piscataway, New Jersey 08855, USA}
\author{M.~Hare$^{\dag}$}~\affiliation{Tufts University, Medford, Massachusetts 02155, USA}
\author{A.~Harel$^{\ddag}$}~\affiliation{University of Rochester, Rochester, New York 14627, USA}
\author{R.F.~Harr$^{\dag}$}~\affiliation{Wayne State University, Detroit, Michigan 48201, USA}
\author{K.~Hatakeyama$^{\dag}$}~\affiliation{Baylor University, Waco, Texas 76798, USA}
\author{J.M.~Hauptman$^{\ddag}$}~\affiliation{Iowa State University, Ames, Iowa 50011, USA}
\author{C.~Hays$^{\dag}$}~\affiliation{University of Oxford, Oxford OX1 3RH, United Kingdom}
\author{J.~Hays$^{\ddag}$}~\affiliation{Imperial College London, London SW7 2AZ, United Kingdom}
\author{T.~Head$^{\ddag}$}~\affiliation{The University of Manchester, Manchester M13 9PL, United Kingdom}
\author{T.~Hebbeker$^{\ddag}$}~\affiliation{III. Physikalisches Institut A, RWTH Aachen University, Aachen, Germany}
\author{M.~Heck$^{\dag}$}~\affiliation{Institut f\"{u}r Experimentelle Kernphysik, Karlsruhe Institute of Technology, D-76131 Karlsruhe, Germany}
\author{D.~Hedin$^{\ddag}$}~\affiliation{Northern Illinois University, DeKalb, Illinois 60115, USA}
\author{H.~Hegab$^{\ddag}$}~\affiliation{Oklahoma State University, Stillwater, Oklahoma 74078, USA}
\author{J.~Heinrich$^{\dag}$}~\affiliation{University of Pennsylvania, Philadelphia, Pennsylvania 19104, USA}
\author{A.P.~Heinson$^{\ddag}$}~\affiliation{University of California Riverside, Riverside, California 92521, USA}
\author{U.~Heintz$^{\ddag}$}~\affiliation{Brown University, Providence, Rhode Island 02912, USA}
\author{C.~Hensel$^{\ddag}$}~\affiliation{II. Physikalisches Institut, Georg-August-Universit\"at G\"ottingen, G\"ottingen, Germany}
\author{I.~Heredia-De~La~Cruz$^{\ddag}$}~\affiliation{CINVESTAV, Mexico City, Mexico}
\author{M.~Herndon$^{\dag}$}~\affiliation{University of Wisconsin, Madison, Wisconsin 53706, USA}
\author{K.~Herner$^{\ddag}$}~\affiliation{University of Michigan, Ann Arbor, Michigan 48109, USA}
\author{G.~Hesketh$^{\ddag e}$}~\affiliation{The University of Manchester, Manchester M13 9PL, United Kingdom}
\author{S.~Hewamanage$^{\dag}$}~\affiliation{Baylor University, Waco, Texas 76798, USA}
\author{M.D.~Hildreth$^{\ddag}$}~\affiliation{University of Notre Dame, Notre Dame, Indiana 46556, USA}
\author{R.~Hirosky$^{\ddag}$}~\affiliation{University of Virginia, Charlottesville, Virginia 22904, USA}
\author{T.~Hoang$^{\ddag}$}~\affiliation{Florida State University, Tallahassee, Florida 32306, USA}
\author{J.D.~Hobbs$^{\ddag}$}~\affiliation{State University of New York, Stony Brook, New York 11794, USA}
\author{A.~Hocker$^{\dag}$}~\affiliation{Fermi National Accelerator Laboratory, Batavia, Illinois 60510, USA}
\author{B.~Hoeneisen$^{\ddag}$}~\affiliation{Universidad San Francisco de Quito, Quito, Ecuador}
\author{J.~Hogan$^{\ddag}$}~\affiliation{Rice University, Houston, Texas 77005, USA}
\author{M.~Hohlfeld$^{\ddag}$}~\affiliation{Institut f\"ur Physik, Universit\"at Mainz, Mainz, Germany}
\author{W.~Hopkins$^{\dag j}$}~\affiliation{Fermi National Accelerator Laboratory, Batavia, Illinois 60510, USA}
\author{D.~Horn$^{\dag}$}~\affiliation{Institut f\"{u}r Experimentelle Kernphysik, Karlsruhe Institute of Technology, D-76131 Karlsruhe, Germany}
\author{S.~Hou$^{\dag}$}~\affiliation{Institute of Physics, Academia Sinica, Taipei, Taiwan 11529, Republic of China}
\author{I.~Howley$^{\ddag}$}~\affiliation{University of Texas, Arlington, Texas 76019, USA}
\author{Z.~Hubacek$^{\ddag}$}~\affiliation{Czech Technical University in Prague, Prague, Czech Republic}~\affiliation{CEA, Irfu, SPP, Saclay, France}
\author{R.E.~Hughes$^{\dag}$}~\affiliation{The Ohio State University, Columbus, Ohio 43210, USA}
\author{M.~Hurwitz$^{\dag}$}~\affiliation{Enrico Fermi Institute, University of Chicago, Chicago, Illinois 60637, USA}
\author{U.~Husemann$^{\dag}$}~\affiliation{Yale University, New Haven, Connecticut 06520, USA}
\author{N.~Hussain$^{\dag}$}~\affiliation{Institute of Particle Physics: McGill University, Montr\'{e}al, Qu\'{e}bec, Canada H3A~2T8; Simon Fraser University, Burnaby, British Columbia, Canada V5A~1S6; University of Toronto, Toronto, Ontario, Canada M5S~1A7; and TRIUMF, Vancouver, British Columbia, Canada V6T~2A3}
\author{M.~Hussein$^{\dag}$}~\affiliation{Michigan State University, East Lansing, Michigan 48824, USA}
\author{J.~Huston$^{\dag}$}~\affiliation{Michigan State University, East Lansing, Michigan 48824, USA}
\author{V.~Hynek$^{\ddag}$}~\affiliation{Czech Technical University in Prague, Prague, Czech Republic}
\author{I.~Iashvili$^{\ddag}$}~\affiliation{State University of New York, Buffalo, New York 14260, USA}
\author{Y.~Ilchenko$^{\ddag}$}~\affiliation{Southern Methodist University, Dallas, Texas 75275, USA}
\author{R.~Illingworth$^{\ddag}$}~\affiliation{Fermi National Accelerator Laboratory, Batavia, Illinois 60510, USA}
\author{G.~Introzzi$^{\dag}$}~\affiliation{Istituto Nazionale di Fisica Nucleare Pisa, $^{\sharp{c}}$University of Pisa, $^{\sharp{d}}$University of Siena and $^{\sharp{e}}$Scuola Normale Superiore, I-56127 Pisa, Italy}
\author{M.~Iori$^{\dag\sharp{f}}$}~\affiliation{Istituto Nazionale di Fisica Nucleare, Sezione di Roma 1, $^{\sharp{f}}$Sapienza Universit\`{a} di Roma, I-00185 Roma, Italy}
\author{A.S.~Ito$^{\ddag}$}~\affiliation{Fermi National Accelerator Laboratory, Batavia, Illinois 60510, USA}
\author{A.~Ivanov$^{\dag k}$}~\affiliation{University of California, Davis, Davis, California 95616, USA}
\author{S.~Jabeen$^{\ddag}$}~\affiliation{Brown University, Providence, Rhode Island 02912, USA}
\author{M.~Jaffr\'e$^{\ddag}$}~\affiliation{LAL, Universit\'e Paris-Sud, CNRS/IN2P3, Orsay, France}
\author{E.~James$^{\dag}$}~\affiliation{Fermi National Accelerator Laboratory, Batavia, Illinois 60510, USA}
\author{D.~Jang$^{\dag}$}~\affiliation{Carnegie Mellon University, Pittsburgh, Pennsylvania 15213, USA}
\author{A.~Jayasinghe$^{\ddag}$}~\affiliation{University of Oklahoma, Norman, Oklahoma 73019, USA}
\author{B.~Jayatilaka$^{\dag}$}~\affiliation{Duke University, Durham, North Carolina 27708, USA}
\author{E.J.~Jeon$^{\dag}$}~\affiliation{Center for High Energy Physics: Kyungpook National University, Daegu 702-701, Korea; Seoul National University, Seoul 151-742, Korea; Sungkyunkwan University, Suwon 440-746, Korea; Korea Institute of Science and Technology Information, Daejeon 305-806, Korea; Chonnam National University, Gwangju 500-757, Korea; Chonbuk National University, Jeonju 561-756, Korea}
\author{M.S.~Jeong$^{\ddag}$}~\affiliation{Korea Detector Laboratory, Korea University, Seoul, Korea}
\author{R.~Jesik$^{\ddag}$}~\affiliation{Imperial College London, London SW7 2AZ, United Kingdom}
\author{S.~Jindariani$^{\dag}$}~\affiliation{Fermi National Accelerator Laboratory, Batavia, Illinois 60510, USA}
\author{K.~Johns$^{\ddag}$}~\affiliation{University of Arizona, Tucson, Arizona 85721, USA}
\author{E.~Johnson$^{\ddag}$}~\affiliation{Michigan State University, East Lansing, Michigan 48824, USA}
\author{M.~Johnson$^{\ddag}$}~\affiliation{Fermi National Accelerator Laboratory, Batavia, Illinois 60510, USA}
\author{A.~Jonckheere$^{\ddag}$}~\affiliation{Fermi National Accelerator Laboratory, Batavia, Illinois 60510, USA}
\author{M.~Jones$^{\dag}$}~\affiliation{Purdue University, West Lafayette, Indiana 47907, USA}
\author{P.~Jonsson$^{\ddag}$}~\affiliation{Imperial College London, London SW7 2AZ, United Kingdom}
\author{K.K.~Joo$^{\dag}$}~\affiliation{Center for High Energy Physics: Kyungpook National University, Daegu 702-701, Korea; Seoul National University, Seoul 151-742, Korea; Sungkyunkwan University, Suwon 440-746, Korea; Korea Institute of Science and Technology Information, Daejeon 305-806, Korea; Chonnam National University, Gwangju 500-757, Korea; Chonbuk National University, Jeonju 561-756, Korea}
\author{J.~Joshi$^{\ddag}$}~\affiliation{University of California Riverside, Riverside, California 92521, USA}
\author{S.Y.~Jun$^{\dag}$}~\affiliation{Carnegie Mellon University, Pittsburgh, Pennsylvania 15213, USA}
\author{A.W.~Jung$^{\ddag}$}~\affiliation{Fermi National Accelerator Laboratory, Batavia, Illinois 60510, USA}
\author{T.R.~Junk$^{\dag}$}~\affiliation{Fermi National Accelerator Laboratory, Batavia, Illinois 60510, USA}
\author{A.~Juste$^{\ddag}$}~\affiliation{Instituci\'{o} Catalana de Recerca i Estudis Avan\c{c}ats (ICREA) and Institut de F\'{i}sica d'Altes Energies (IFAE), Barcelona, Spain}
\author{K.~Kaadze$^{\ddag}$}~\affiliation{Kansas State University, Manhattan, Kansas 66506, USA}
\author{E.~Kajfasz$^{\ddag}$}~\affiliation{CPPM, Aix-Marseille Universit\'e, CNRS/IN2P3, Marseille, France}
\author{T.~Kamon$^{\dag}$}~\affiliation{Texas A\&M University, College Station, Texas 77843, USA}
\author{P.E.~Karchin$^{\dag}$}~\affiliation{Wayne State University, Detroit, Michigan 48201, USA}
\author{D.~Karmanov$^{\ddag}$}~\affiliation{Moscow State University, Moscow, Russia}
\author{A.~Kasmi$^{\dag}$}~\affiliation{Baylor University, Waco, Texas 76798, USA}
\author{P.A.~Kasper$^{\ddag}$}~\affiliation{Fermi National Accelerator Laboratory, Batavia, Illinois 60510, USA}
\author{Y.~Kato$^{\dag l}$}~\affiliation{Osaka City University, Osaka 588, Japan}
\author{I.~Katsanos$^{\ddag}$}~\affiliation{University of Nebraska, Lincoln, Nebraska 68588, USA}
\author{R.~Kehoe$^{\ddag}$}~\affiliation{Southern Methodist University, Dallas, Texas 75275, USA}
\author{S.~Kermiche$^{\ddag}$}~\affiliation{CPPM, Aix-Marseille Universit\'e, CNRS/IN2P3, Marseille, France}
\author{W.~Ketchum$^{\dag}$}~\affiliation{Enrico Fermi Institute, University of Chicago, Chicago, Illinois 60637, USA}
\author{J.~Keung$^{\dag}$}~\affiliation{University of Pennsylvania, Philadelphia, Pennsylvania 19104, USA}
\author{N.~Khalatyan$^{\ddag}$}~\affiliation{Fermi National Accelerator Laboratory, Batavia, Illinois 60510, USA}
\author{A.~Khanov$^{\ddag}$}~\affiliation{Oklahoma State University, Stillwater, Oklahoma 74078, USA}
\author{A.~Kharchilava$^{\ddag}$}~\affiliation{State University of New York, Buffalo, New York 14260, USA}
\author{Y.N.~Kharzheev$^{\ddag}$}~\affiliation{Joint Institute for Nuclear Research, Dubna, Russia}
\author{V.~Khotilovich$^{\dag}$}~\affiliation{Texas A\&M University, College Station, Texas 77843, USA}
\author{B.~Kilminster$^{\dag}$}~\affiliation{Fermi National Accelerator Laboratory, Batavia, Illinois 60510, USA}
\author{D.H.~Kim$^{\dag}$}~\affiliation{Center for High Energy Physics: Kyungpook National University, Daegu 702-701, Korea; Seoul National University, Seoul 151-742, Korea; Sungkyunkwan University, Suwon 440-746, Korea; Korea Institute of Science and Technology Information, Daejeon 305-806, Korea; Chonnam National University, Gwangju 500-757, Korea; Chonbuk National University, Jeonju 561-756, Korea}
\author{H.S.~Kim$^{\dag}$}~\affiliation{Center for High Energy Physics: Kyungpook National University, Daegu 702-701, Korea; Seoul National University, Seoul 151-742, Korea; Sungkyunkwan University, Suwon 440-746, Korea; Korea Institute of Science and Technology Information, Daejeon 305-806, Korea; Chonnam National University, Gwangju 500-757, Korea; Chonbuk National University, Jeonju 561-756, Korea}
\author{J.E.~Kim$^{\dag}$}~\affiliation{Center for High Energy Physics: Kyungpook National University, Daegu 702-701, Korea; Seoul National University, Seoul 151-742, Korea; Sungkyunkwan University, Suwon 440-746, Korea; Korea Institute of Science and Technology Information, Daejeon 305-806, Korea; Chonnam National University, Gwangju 500-757, Korea; Chonbuk National University, Jeonju 561-756, Korea}
\author{M.J.~Kim$^{\dag}$}~\affiliation{Laboratori Nazionali di Frascati, Istituto Nazionale di Fisica Nucleare, I-00044 Frascati, Italy}
\author{S.B.~Kim$^{\dag}$}~\affiliation{Center for High Energy Physics: Kyungpook National University, Daegu 702-701, Korea; Seoul National University, Seoul 151-742, Korea; Sungkyunkwan University, Suwon 440-746, Korea; Korea Institute of Science and Technology Information, Daejeon 305-806, Korea; Chonnam National University, Gwangju 500-757, Korea; Chonbuk National University, Jeonju 561-756, Korea}
\author{S.H.~Kim$^{\dag}$}~\affiliation{University of Tsukuba, Tsukuba, Ibaraki 305, Japan}
\author{Y.J.~Kim$^{\dag}$}~\affiliation{Center for High Energy Physics: Kyungpook National University, Daegu 702-701, Korea; Seoul National University, Seoul 151-742, Korea; Sungkyunkwan University, Suwon 440-746, Korea; Korea Institute of Science and Technology Information, Daejeon 305-806, Korea; Chonnam National University, Gwangju 500-757, Korea; Chonbuk National University, Jeonju 561-756, Korea}
\author{Y.K.~Kim$^{\dag}$}~\affiliation{Enrico Fermi Institute, University of Chicago, Chicago, Illinois 60637, USA}
\author{N.~Kimura$^{\dag}$}~\affiliation{Waseda University, Tokyo 169, Japan}
\author{M.~Kirby$^{\dag}$}~\affiliation{Fermi National Accelerator Laboratory, Batavia, Illinois 60510, USA}
\author{I.~Kiselevich$^{\ddag}$}~\affiliation{Institute for Theoretical and Experimental Physics, Moscow, Russia}
\author{S.~Klimenko$^{\dag}$}~\affiliation{University of Florida, Gainesville, Florida 32611, USA}
\author{K.~Knoepfel$^{\dag}$}~\affiliation{Fermi National Accelerator Laboratory, Batavia, Illinois 60510, USA}
\author{J.M.~Kohli$^{\ddag}$}~\affiliation{Panjab University, Chandigarh, India}
\author{K.~Kondo\footnote{Deceased}$^{\dag}$}~\affiliation{Waseda University, Tokyo 169, Japan}
\author{D.J.~Kong$^{\dag}$}~\affiliation{Center for High Energy Physics: Kyungpook National University, Daegu 702-701, Korea; Seoul National University, Seoul 151-742, Korea; Sungkyunkwan University, Suwon 440-746, Korea; Korea Institute of Science and Technology Information, Daejeon 305-806, Korea; Chonnam National University, Gwangju 500-757, Korea; Chonbuk National University, Jeonju 561-756, Korea}
\author{J.~Konigsberg$^{\dag}$}~\affiliation{University of Florida, Gainesville, Florida 32611, USA}
\author{A.V.~Kotwal$^{\dag}$}~\affiliation{Duke University, Durham, North Carolina 27708, USA}
\author{A.V.~Kozelov$^{\ddag}$}~\affiliation{Institute for High Energy Physics, Protvino, Russia}
\author{J.~Kraus$^{\ddag}$}~\affiliation{University of Mississippi, University, Mississippi 38677, USA}
\author{M.~Kreps$^{\dag}$}~\affiliation{Institut f\"{u}r Experimentelle Kernphysik, Karlsruhe Institute of Technology, D-76131 Karlsruhe, Germany}
\author{J.~Kroll$^{\dag}$}~\affiliation{University of Pennsylvania, Philadelphia, Pennsylvania 19104, USA}
\author{D.~Krop$^{\dag}$}~\affiliation{Enrico Fermi Institute, University of Chicago, Chicago, Illinois 60637, USA}
\author{M.~Kruse$^{\dag}$}~\affiliation{Duke University, Durham, North Carolina 27708, USA}
\author{V.~Krutelyov$^{\dag m}$}~\affiliation{Texas A\&M University, College Station, Texas 77843, USA}
\author{T.~Kuhr$^{\dag}$}~\affiliation{Institut f\"{u}r Experimentelle Kernphysik, Karlsruhe Institute of Technology, D-76131 Karlsruhe, Germany}
\author{S.~Kulikov$^{\ddag}$}~\affiliation{Institute for High Energy Physics, Protvino, Russia}
\author{A.~Kumar$^{\ddag}$}~\affiliation{State University of New York, Buffalo, New York 14260, USA}
\author{A.~Kupco$^{\ddag}$}~\affiliation{Center for Particle Physics, Institute of Physics, Academy of Sciences of the Czech Republic, Prague, Czech Republic}
\author{M.~Kurata$^{\dag}$}~\affiliation{University of Tsukuba, Tsukuba, Ibaraki 305, Japan}
\author{T.~Kur\v{c}a$^{\ddag}$}~\affiliation{IPNL, Universit\'e Lyon 1, CNRS/IN2P3, Villeurbanne, France and Universit\'e de Lyon, Lyon, France}
\author{V.A.~Kuzmin$^{\ddag}$}~\affiliation{Moscow State University, Moscow, Russia}
\author{S.~Kwang$^{\dag}$}~\affiliation{Enrico Fermi Institute, University of Chicago, Chicago, Illinois 60637, USA}
\author{A.T.~Laasanen$^{\dag}$}~\affiliation{Purdue University, West Lafayette, Indiana 47907, USA}
\author{S.~Lami$^{\dag}$}~\affiliation{Istituto Nazionale di Fisica Nucleare Pisa, $^{\sharp{c}}$University of Pisa, $^{\sharp{d}}$University of Siena and $^{\sharp{e}}$Scuola Normale Superiore, I-56127 Pisa, Italy}
\author{S.~Lammel$^{\dag}$}~\affiliation{Fermi National Accelerator Laboratory, Batavia, Illinois 60510, USA}
\author{S.~Lammers$^{\ddag}$}~\affiliation{Indiana University, Bloomington, Indiana 47405, USA}
\author{M.~Lancaster$^{\dag}$}~\affiliation{University College London, London WC1E 6BT, United Kingdom}
\author{R.L.~Lander$^{\dag}$}~\affiliation{University of California, Davis, Davis, California 95616, USA}
\author{G.~Landsberg$^{\ddag}$}~\affiliation{Brown University, Providence, Rhode Island 02912, USA}
\author{K.~Lannon$^{\dag n}$}~\affiliation{The Ohio State University, Columbus, Ohio 43210, USA}
\author{A.~Lath$^{\dag}$}~\affiliation{Rutgers University, Piscataway, New Jersey 08855, USA}
\author{G.~Latino$^{\dag\sharp{d}}$}~\affiliation{Istituto Nazionale di Fisica Nucleare Pisa, $^{\sharp{c}}$University of Pisa, $^{\sharp{d}}$University of Siena and $^{\sharp{e}}$Scuola Normale Superiore, I-56127 Pisa, Italy}
\author{P.~Lebrun$^{\ddag}$}~\affiliation{IPNL, Universit\'e Lyon 1, CNRS/IN2P3, Villeurbanne, France and Universit\'e de Lyon, Lyon, France}
\author{T.~LeCompte$^{\dag}$}~\affiliation{Argonne National Laboratory, Argonne, Illinois 60439, USA}
\author{E.~Lee$^{\dag}$}~\affiliation{Texas A\&M University, College Station, Texas 77843, USA}
\author{H.S.~Lee$^{\ddag}$}~\affiliation{Korea Detector Laboratory, Korea University, Seoul, Korea}
\author{H.S.~Lee$^{\dag o}$}~\affiliation{Enrico Fermi Institute, University of Chicago, Chicago, Illinois 60637, USA}
\author{J.S.~Lee$^{\dag}$}~\affiliation{Center for High Energy Physics: Kyungpook National University, Daegu 702-701, Korea; Seoul National University, Seoul 151-742, Korea; Sungkyunkwan University, Suwon 440-746, Korea; Korea Institute of Science and Technology Information, Daejeon 305-806, Korea; Chonnam National University, Gwangju 500-757, Korea; Chonbuk National University, Jeonju 561-756, Korea}
\author{S.W.~Lee$^{\dag p}$}~\affiliation{Texas A\&M University, College Station, Texas 77843, USA}
\author{S.W.~Lee$^{\ddag}$}~\affiliation{Iowa State University, Ames, Iowa 50011, USA}
\author{W.M.~Lee$^{\ddag}$}~\affiliation{Fermi National Accelerator Laboratory, Batavia, Illinois 60510, USA}
\author{X.~Lei$^{\ddag}$}~\affiliation{University of Arizona, Tucson, Arizona 85721, USA}
\author{J.~Lellouch$^{\ddag}$}~\affiliation{LPNHE, Universit\'es Paris VI and VII, CNRS/IN2P3, Paris, France}
\author{S.~Leo$^{\dag\sharp{c}}$}~\affiliation{Istituto Nazionale di Fisica Nucleare Pisa, $^{\sharp{c}}$University of Pisa, $^{\sharp{d}}$University of Siena and $^{\sharp{e}}$Scuola Normale Superiore, I-56127 Pisa, Italy}
\author{S.~Leone$^{\dag}$}~\affiliation{Istituto Nazionale di Fisica Nucleare Pisa, $^{\sharp{c}}$University of Pisa, $^{\sharp{d}}$University of Siena and $^{\sharp{e}}$Scuola Normale Superiore, I-56127 Pisa, Italy}
\author{J.D.~Lewis$^{\dag}$}~\affiliation{Fermi National Accelerator Laboratory, Batavia, Illinois 60510, USA}
\author{H.~Li$^{\ddag}$}~\affiliation{LPSC, Universit\'e Joseph Fourier Grenoble 1, CNRS/IN2P3, Institut National Polytechnique de Grenoble, Grenoble, France}
\author{L.~Li$^{\ddag}$}~\affiliation{University of California Riverside, Riverside, California 92521, USA}
\author{Q.Z.~Li$^{\ddag}$}~\affiliation{Fermi National Accelerator Laboratory, Batavia, Illinois 60510, USA}
\author{J.K.~Lim$^{\ddag}$}~\affiliation{Korea Detector Laboratory, Korea University, Seoul, Korea}
\author{A.~Limosani$^{\dag q}$}~\affiliation{Duke University, Durham, North Carolina 27708, USA}
\author{D.~Lincoln$^{\ddag}$}~\affiliation{Fermi National Accelerator Laboratory, Batavia, Illinois 60510, USA}
\author{C.-J.~Lin$^{\dag}$}~\affiliation{Ernest Orlando Lawrence Berkeley National Laboratory, Berkeley, California 94720, USA}
\author{M.~Lindgren$^{\dag}$}~\affiliation{Fermi National Accelerator Laboratory, Batavia, Illinois 60510, USA}
\author{J.~Linnemann$^{\ddag}$}~\affiliation{Michigan State University, East Lansing, Michigan 48824, USA}
\author{V.V.~Lipaev$^{\ddag}$}~\affiliation{Institute for High Energy Physics, Protvino, Russia}
\author{E.~Lipeles$^{\dag}$}~\affiliation{University of Pennsylvania, Philadelphia, Pennsylvania 19104, USA}
\author{R.~Lipton$^{\ddag}$}~\affiliation{Fermi National Accelerator Laboratory, Batavia, Illinois 60510, USA}
\author{A.~Lister$^{\dag}$}~\affiliation{University of Geneva, CH-1211 Geneva 4, Switzerland}
\author{D.O.~Litvintsev$^{\dag}$}~\affiliation{Fermi National Accelerator Laboratory, Batavia, Illinois 60510, USA}
\author{C.~Liu$^{\dag}$}~\affiliation{University of Pittsburgh, Pittsburgh, Pennsylvania 15260, USA}
\author{H.~Liu$^{\dag}$}~\affiliation{University of Virginia, Charlottesville, Virginia 22906, USA}
\author{H.~Liu$^{\ddag}$}~\affiliation{Southern Methodist University, Dallas, Texas 75275, USA}
\author{Q.~Liu$^{\dag}$}~\affiliation{Purdue University, West Lafayette, Indiana 47907, USA}
\author{T.~Liu$^{\dag}$}~\affiliation{Fermi National Accelerator Laboratory, Batavia, Illinois 60510, USA}
\author{Y.~Liu$^{\ddag}$}~\affiliation{University of Science and Technology of China, Hefei, People's Republic of China}
\author{A.~Lobodenko$^{\ddag}$}~\affiliation{Petersburg Nuclear Physics Institute, St. Petersburg, Russia}
\author{S.~Lockwitz$^{\dag}$}~\affiliation{Yale University, New Haven, Connecticut 06520, USA}
\author{A.~Loginov$^{\dag}$}~\affiliation{Yale University, New Haven, Connecticut 06520, USA}
\author{M.~Lokajicek$^{\ddag}$}~\affiliation{Center for Particle Physics, Institute of Physics, Academy of Sciences of the Czech Republic, Prague, Czech Republic}
\author{R.~Lopes~de~Sa$^{\ddag}$}~\affiliation{State University of New York, Stony Brook, New York 11794, USA}
\author{H.J.~Lubatti$^{\ddag}$}~\affiliation{University of Washington, Seattle, Washington 98195, USA}
\author{D.~Lucchesi$^{\dag\sharp{b}}$}~\affiliation{Istituto Nazionale di Fisica Nucleare, Sezione di Padova-Trento, $^{\sharp{b}}$University of Padova, I-35131 Padova, Italy}
\author{J.~Lueck$^{\dag}$}~\affiliation{Institut f\"{u}r Experimentelle Kernphysik, Karlsruhe Institute of Technology, D-76131 Karlsruhe, Germany}
\author{P.~Lujan$^{\dag}$}~\affiliation{Ernest Orlando Lawrence Berkeley National Laboratory, Berkeley, California 94720, USA}
\author{P.~Lukens$^{\dag}$}~\affiliation{Fermi National Accelerator Laboratory, Batavia, Illinois 60510, USA}
\author{R.~Luna-Garcia$^{\ddag f}$}~\affiliation{CINVESTAV, Mexico City, Mexico}
\author{G.~Lungu$^{\dag}$}~\affiliation{The Rockefeller University, New York, New York 10065, USA}
\author{A.L.~Lyon$^{\ddag}$}~\affiliation{Fermi National Accelerator Laboratory, Batavia, Illinois 60510, USA}
\author{R.~Lysak$^{\dag r}$}~\affiliation{Comenius University, 842 48 Bratislava, Slovakia; Institute of Experimental Physics, 040 01 Kosice, Slovakia}
\author{J.~Lys$^{\dag}$}~\affiliation{Ernest Orlando Lawrence Berkeley National Laboratory, Berkeley, California 94720, USA}
\author{A.K.A.~Maciel$^{\ddag}$}~\affiliation{LAFEX, Centro Brasileiro de Pesquisas F\'{i}sicas, Rio de Janeiro, Brazil}
\author{R.~Madar$^{\ddag}$}~\affiliation{CEA, Irfu, SPP, Saclay, France}
\author{R.~Madrak$^{\dag}$}~\affiliation{Fermi National Accelerator Laboratory, Batavia, Illinois 60510, USA}
\author{K.~Maeshima$^{\dag}$}~\affiliation{Fermi National Accelerator Laboratory, Batavia, Illinois 60510, USA}
\author{P.~Maestro$^{\dag\sharp{d}}$}~\affiliation{Istituto Nazionale di Fisica Nucleare Pisa, $^{\sharp{c}}$University of Pisa, $^{\sharp{d}}$University of Siena and $^{\sharp{e}}$Scuola Normale Superiore, I-56127 Pisa, Italy}
\author{R.~Maga\~na-Villalba$^{\ddag}$}~\affiliation{CINVESTAV, Mexico City, Mexico}
\author{S.~Malik$^{\dag}$}~\affiliation{The Rockefeller University, New York, New York 10065, USA}
\author{S.~Malik$^{\ddag}$}~\affiliation{University of Nebraska, Lincoln, Nebraska 68588, USA}
\author{V.L.~Malyshev$^{\ddag}$}~\affiliation{Joint Institute for Nuclear Research, Dubna, Russia}
\author{G.~Manca$^{\dag s}$}~\affiliation{University of Liverpool, Liverpool L69 7ZE, United Kingdom}
\author{A.~Manousakis-Katsikakis$^{\dag}$}~\affiliation{University of Athens, 157 71 Athens, Greece}
\author{Y.~Maravin$^{\ddag}$}~\affiliation{Kansas State University, Manhattan, Kansas 66506, USA}
\author{F.~Margaroli$^{\dag}$}~\affiliation{Istituto Nazionale di Fisica Nucleare, Sezione di Roma 1, $^{\sharp{f}}$Sapienza Universit\`{a} di Roma, I-00185 Roma, Italy}
\author{C.~Marino$^{\dag}$}~\affiliation{Institut f\"{u}r Experimentelle Kernphysik, Karlsruhe Institute of Technology, D-76131 Karlsruhe, Germany}
\author{M.~Mart\'{\i}nez$^{\dag}$}~\affiliation{Institut de Fisica d'Altes Energies, ICREA, Universitat Autonoma de Barcelona, E-08193, Bellaterra (Barcelona), Spain}
\author{J.~Mart\'{\i}nez-Ortega$^{\ddag}$}~\affiliation{CINVESTAV, Mexico City, Mexico}
\author{P.~Mastrandrea$^{\dag}$}~\affiliation{Istituto Nazionale di Fisica Nucleare, Sezione di Roma 1, $^{\sharp{f}}$Sapienza Universit\`{a} di Roma, I-00185 Roma, Italy}
\author{K.~Matera$^{\dag}$}~\affiliation{University of Illinois, Urbana, Illinois 61801, USA}
\author{M.E.~Mattson$^{\dag}$}~\affiliation{Wayne State University, Detroit, Michigan 48201, USA}
\author{A.~Mazzacane$^{\dag}$}~\affiliation{Fermi National Accelerator Laboratory, Batavia, Illinois 60510, USA}
\author{P.~Mazzanti$^{\dag}$}~\affiliation{Istituto Nazionale di Fisica Nucleare Bologna, $^{\sharp{a}}$University of Bologna, I-40127 Bologna, Italy}
\author{R.~McCarthy$^{\ddag}$}~\affiliation{State University of New York, Stony Brook, New York 11794, USA}
\author{K.S.~McFarland$^{\dag}$}~\affiliation{University of Rochester, Rochester, New York 14627, USA}
\author{C.L.~McGivern$^{\ddag}$}~\affiliation{The University of Manchester, Manchester M13 9PL, United Kingdom}
\author{P.~McIntyre$^{\dag}$}~\affiliation{Texas A\&M University, College Station, Texas 77843, USA}
\author{R.~McNulty$^{\dag t}$}~\affiliation{University of Liverpool, Liverpool L69 7ZE, United Kingdom}
\author{A.~Mehta$^{\dag}$}~\affiliation{University of Liverpool, Liverpool L69 7ZE, United Kingdom}
\author{P.~Mehtala$^{\dag}$}~\affiliation{Division of High Energy Physics, Department of Physics, University of Helsinki and Helsinki Institute of Physics, FIN-00014, Helsinki, Finland}
\author{M.M.~Meijer$^{\ddag}$}~\affiliation{Nikhef, Science Park, Amsterdam, the Netherlands}~\affiliation{Radboud University Nijmegen, Nijmegen, the Netherlands}
\author{A.~Melnitchouk$^{\ddag}$}~\affiliation{University of Mississippi, University, Mississippi 38677, USA}
\author{D.~Menezes$^{\ddag}$}~\affiliation{Northern Illinois University, DeKalb, Illinois 60115, USA}
\author{P.G.~Mercadante$^{\ddag}$}~\affiliation{Universidade Federal do ABC, Santo Andr\'e, Brazil}
\author{M.~Merkin$^{\ddag}$}~\affiliation{Moscow State University, Moscow, Russia}
\author{A.~Meyer$^{\ddag}$}~\affiliation{III. Physikalisches Institut A, RWTH Aachen University, Aachen, Germany}
\author{J.~Meyer$^{\ddag}$}~\affiliation{II. Physikalisches Institut, Georg-August-Universit\"at G\"ottingen, G\"ottingen, Germany}
\author{T.~Miao$^{\dag}$}~\affiliation{Fermi National Accelerator Laboratory, Batavia, Illinois 60510, USA}
\author{F.~Miconi$^{\ddag}$}~\affiliation{IPHC, Universit\'e de Strasbourg, CNRS/IN2P3, Strasbourg, France}
\author{D.~Mietlicki$^{\dag}$}~\affiliation{University of Michigan, Ann Arbor, Michigan 48109, USA}
\author{A.~Mitra$^{\dag}$}~\affiliation{Institute of Physics, Academia Sinica, Taipei, Taiwan 11529, Republic of China}
\author{H.~Miyake$^{\dag}$}~\affiliation{University of Tsukuba, Tsukuba, Ibaraki 305, Japan}
\author{S.~Moed$^{\dag}$}~\affiliation{Fermi National Accelerator Laboratory, Batavia, Illinois 60510, USA}
\author{N.~Moggi$^{\dag}$}~\affiliation{Istituto Nazionale di Fisica Nucleare Bologna, $^{\sharp{a}}$University of Bologna, I-40127 Bologna, Italy}
\author{N.K.~Mondal$^{\ddag}$}~\affiliation{Tata Institute of Fundamental Research, Mumbai, India}
\author{M.N.~Mondragon$^{\dag c}$}~\affiliation{Fermi National Accelerator Laboratory, Batavia, Illinois 60510, USA}
\author{C.S.~Moon$^{\dag}$}~\affiliation{Center for High Energy Physics: Kyungpook National University, Daegu 702-701, Korea; Seoul National University, Seoul 151-742, Korea; Sungkyunkwan University, Suwon 440-746, Korea; Korea Institute of Science and Technology Information, Daejeon 305-806, Korea; Chonnam National University, Gwangju 500-757, Korea; Chonbuk National University, Jeonju 561-756, Korea}
\author{R.~Moore$^{\dag}$}~\affiliation{Fermi National Accelerator Laboratory, Batavia, Illinois 60510, USA}
\author{M.J.~Morello$^{\dag\sharp{e}}$}~\affiliation{Istituto Nazionale di Fisica Nucleare Pisa, $^{\sharp{c}}$University of Pisa, $^{\sharp{d}}$University of Siena and $^{\sharp{e}}$Scuola Normale Superiore, I-56127 Pisa, Italy}
\author{J.~Morlock$^{\dag}$}~\affiliation{Institut f\"{u}r Experimentelle Kernphysik, Karlsruhe Institute of Technology, D-76131 Karlsruhe, Germany}
\author{P.~Movilla~Fernandez$^{\dag}$}~\affiliation{Fermi National Accelerator Laboratory, Batavia, Illinois 60510, USA}
\author{A.~Mukherjee$^{\dag}$}~\affiliation{Fermi National Accelerator Laboratory, Batavia, Illinois 60510, USA}
\author{M.~Mulhearn$^{\ddag}$}~\affiliation{University of Virginia, Charlottesville, Virginia 22904, USA}
\author{Th.~Muller$^{\dag}$}~\affiliation{Institut f\"{u}r Experimentelle Kernphysik, Karlsruhe Institute of Technology, D-76131 Karlsruhe, Germany}
\author{P.~Murat$^{\dag}$}~\affiliation{Fermi National Accelerator Laboratory, Batavia, Illinois 60510, USA}
\author{M.~Mussini$^{\dag\sharp{a}}$}~\affiliation{Istituto Nazionale di Fisica Nucleare Bologna, $^{\sharp{a}}$University of Bologna, I-40127 Bologna, Italy}
\author{J.~Nachtman$^{\dag u}$}~\affiliation{Fermi National Accelerator Laboratory, Batavia, Illinois 60510, USA}
\author{Y.~Nagai$^{\dag}$}~\affiliation{University of Tsukuba, Tsukuba, Ibaraki 305, Japan}
\author{J.~Naganoma$^{\dag}$}~\affiliation{Waseda University, Tokyo 169, Japan}
\author{E.~Nagy$^{\ddag}$}~\affiliation{CPPM, Aix-Marseille Universit\'e, CNRS/IN2P3, Marseille, France}
\author{M.~Naimuddin$^{\ddag}$}~\affiliation{Delhi University, Delhi, India}
\author{I.~Nakano$^{\dag}$}~\affiliation{Okayama University, Okayama 700-8530, Japan}
\author{A.~Napier$^{\dag}$}~\affiliation{Tufts University, Medford, Massachusetts 02155, USA}
\author{M.~Narain$^{\ddag}$}~\affiliation{Brown University, Providence, Rhode Island 02912, USA}
\author{R.~Nayyar$^{\ddag}$}~\affiliation{University of Arizona, Tucson, Arizona 85721, USA}
\author{H.A.~Neal$^{\ddag}$}~\affiliation{University of Michigan, Ann Arbor, Michigan 48109, USA}
\author{J.P.~Negret$^{\ddag}$}~\affiliation{Universidad de los Andes, Bogot\'a, Colombia}
\author{J.~Nett$^{\dag}$}~\affiliation{Texas A\&M University, College Station, Texas 77843, USA}
\author{M.S.~Neubauer$^{\dag}$}~\affiliation{University of Illinois, Urbana, Illinois 61801, USA}
\author{C.~Neu$^{\dag}$}~\affiliation{University of Virginia, Charlottesville, Virginia 22906, USA}
\author{P.~Neustroev$^{\ddag}$}~\affiliation{Petersburg Nuclear Physics Institute, St. Petersburg, Russia}
\author{J.~Nielsen$^{\dag v}$}~\affiliation{Ernest Orlando Lawrence Berkeley National Laboratory, Berkeley, California 94720, USA}
\author{L.~Nodulman$^{\dag}$}~\affiliation{Argonne National Laboratory, Argonne, Illinois 60439, USA}
\author{S.Y.~Noh$^{\dag}$}~\affiliation{Center for High Energy Physics: Kyungpook National University, Daegu 702-701, Korea; Seoul National University, Seoul 151-742, Korea; Sungkyunkwan University, Suwon 440-746, Korea; Korea Institute of Science and Technology Information, Daejeon 305-806, Korea; Chonnam National University, Gwangju 500-757, Korea; Chonbuk National University, Jeonju 561-756, Korea}
\author{O.~Norniella$^{\dag}$}~\affiliation{University of Illinois, Urbana, Illinois 61801, USA}
\author{T.~Nunnemann$^{\ddag}$}~\affiliation{Ludwig-Maximilians-Universit\"at M\"unchen, M\"unchen, Germany}
\author{L.~Oakes$^{\dag}$}~\affiliation{University of Oxford, Oxford OX1 3RH, United Kingdom}
\author{S.H.~Oh$^{\dag}$}~\affiliation{Duke University, Durham, North Carolina 27708, USA}
\author{Y.D.~Oh$^{\dag}$}~\affiliation{Center for High Energy Physics: Kyungpook National University, Daegu 702-701, Korea; Seoul National University, Seoul 151-742, Korea; Sungkyunkwan University, Suwon 440-746, Korea; Korea Institute of Science and Technology Information, Daejeon 305-806, Korea; Chonnam National University, Gwangju 500-757, Korea; Chonbuk National University, Jeonju 561-756, Korea}
\author{I.~Oksuzian$^{\dag}$}~\affiliation{University of Virginia, Charlottesville, Virginia 22906, USA}
\author{T.~Okusawa$^{\dag}$}~\affiliation{Osaka City University, Osaka 588, Japan}
\author{R.~Orava$^{\dag}$}~\affiliation{Division of High Energy Physics, Department of Physics, University of Helsinki and Helsinki Institute of Physics, FIN-00014, Helsinki, Finland}
\author{J.~Orduna$^{\ddag}$}~\affiliation{Rice University, Houston, Texas 77005, USA}
\author{L.~Ortolan$^{\dag}$}~\affiliation{Institut de Fisica d'Altes Energies, ICREA, Universitat Autonoma de Barcelona, E-08193, Bellaterra (Barcelona), Spain}
\author{N.~Osman$^{\ddag}$}~\affiliation{CPPM, Aix-Marseille Universit\'e, CNRS/IN2P3, Marseille, France}
\author{J.~Osta$^{\ddag}$}~\affiliation{University of Notre Dame, Notre Dame, Indiana 46556, USA}
\author{M.~Padilla$^{\ddag}$}~\affiliation{University of California Riverside, Riverside, California 92521, USA}
\author{S.~Pagan~Griso$^{\dag\sharp{b}}$}~\affiliation{Istituto Nazionale di Fisica Nucleare, Sezione di Padova-Trento, $^{\sharp{b}}$University of Padova, I-35131 Padova, Italy}
\author{C.~Pagliarone$^{\dag}$}~\affiliation{Istituto Nazionale di Fisica Nucleare Trieste/Udine, I-34100 Trieste, $^{\sharp{g}}$University of Udine, I-33100 Udine, Italy}
\author{A.~Pal$^{\ddag}$}~\affiliation{University of Texas, Arlington, Texas 76019, USA}
\author{E.~Palencia$^{\dag e}$}~\affiliation{Instituto de Fisica de Cantabria, CSIC-University of Cantabria, 39005 Santander, Spain}
\author{V.~Papadimitriou$^{\dag}$}~\affiliation{Fermi National Accelerator Laboratory, Batavia, Illinois 60510, USA}
\author{A.A.~Paramonov$^{\dag}$}~\affiliation{Argonne National Laboratory, Argonne, Illinois 60439, USA}
\author{N.~Parashar$^{\ddag}$}~\affiliation{Purdue University Calumet, Hammond, Indiana 46323, USA}
\author{V.~Parihar$^{\ddag}$}~\affiliation{Brown University, Providence, Rhode Island 02912, USA}
\author{S.K.~Park$^{\ddag}$}~\affiliation{Korea Detector Laboratory, Korea University, Seoul, Korea}
\author{R.~Partridge$^{\ddag g}$}~\affiliation{Brown University, Providence, Rhode Island 02912, USA}
\author{N.~Parua$^{\ddag}$}~\affiliation{Indiana University, Bloomington, Indiana 47405, USA}
\author{J.~Patrick$^{\dag}$}~\affiliation{Fermi National Accelerator Laboratory, Batavia, Illinois 60510, USA}
\author{A.~Patwa$^{\ddag}$}~\affiliation{Brookhaven National Laboratory, Upton, New York 11973, USA}
\author{G.~Pauletta$^{\dag\sharp{g}}$}~\affiliation{Istituto Nazionale di Fisica Nucleare Trieste/Udine, I-34100 Trieste, $^{\sharp{g}}$University of Udine, I-33100 Udine, Italy}
\author{M.~Paulini$^{\dag}$}~\affiliation{Carnegie Mellon University, Pittsburgh, Pennsylvania 15213, USA}
\author{C.~Paus$^{\dag}$}~\affiliation{Massachusetts Institute of Technology, Cambridge, Massachusetts 02139, USA}
\author{D.E.~Pellett$^{\dag}$}~\affiliation{University of California, Davis, Davis, California 95616, USA}
\author{B.~Penning$^{\ddag}$}~\affiliation{Fermi National Accelerator Laboratory, Batavia, Illinois 60510, USA}
\author{A.~Penzo$^{\dag}$}~\affiliation{Istituto Nazionale di Fisica Nucleare Trieste/Udine, I-34100 Trieste, $^{\sharp{g}}$University of Udine, I-33100 Udine, Italy}
\author{M.~Perfilov$^{\ddag}$}~\affiliation{Moscow State University, Moscow, Russia}
\author{Y.~Peters$^{\ddag}$}~\affiliation{The University of Manchester, Manchester M13 9PL, United Kingdom}
\author{K.~Petridis$^{\ddag}$}~\affiliation{The University of Manchester, Manchester M13 9PL, United Kingdom}
\author{G.~Petrillo$^{\ddag}$}~\affiliation{University of Rochester, Rochester, New York 14627, USA}
\author{P.~P\'etroff$^{\ddag}$}~\affiliation{LAL, Universit\'e Paris-Sud, CNRS/IN2P3, Orsay, France}
\author{T.J.~Phillips$^{\dag}$}~\affiliation{Duke University, Durham, North Carolina 27708, USA}
\author{G.~Piacentino$^{\dag}$}~\affiliation{Istituto Nazionale di Fisica Nucleare Pisa, $^{\sharp{c}}$University of Pisa, $^{\sharp{d}}$University of Siena and $^{\sharp{e}}$Scuola Normale Superiore, I-56127 Pisa, Italy}
\author{E.~Pianori$^{\dag}$}~\affiliation{University of Pennsylvania, Philadelphia, Pennsylvania 19104, USA}
\author{J.~Pilot$^{\dag}$}~\affiliation{The Ohio State University, Columbus, Ohio 43210, USA}
\author{K.~Pitts$^{\dag}$}~\affiliation{University of Illinois, Urbana, Illinois 61801, USA}
\author{C.~Plager$^{\dag}$}~\affiliation{University of California, Los Angeles, Los Angeles, California 90024, USA}
\author{M.-A.~Pleier$^{\ddag}$}~\affiliation{Brookhaven National Laboratory, Upton, New York 11973, USA}
\author{P.L.M.~Podesta-Lerma$^{\ddag h}$}~\affiliation{CINVESTAV, Mexico City, Mexico}
\author{V.M.~Podstavkov$^{\ddag}$}~\affiliation{Fermi National Accelerator Laboratory, Batavia, Illinois 60510, USA}
\author{L.~Pondrom$^{\dag}$}~\affiliation{University of Wisconsin, Madison, Wisconsin 53706, USA}
\author{A.V.~Popov$^{\ddag}$}~\affiliation{Institute for High Energy Physics, Protvino, Russia}
\author{S.~Poprocki$^{\dag j}$}~\affiliation{Fermi National Accelerator Laboratory, Batavia, Illinois 60510, USA}
\author{K.~Potamianos$^{\dag}$}~\affiliation{Purdue University, West Lafayette, Indiana 47907, USA}
\author{A.~Pranko$^{\dag}$}~\affiliation{Ernest Orlando Lawrence Berkeley National Laboratory, Berkeley, California 94720, USA}
\author{M.~Prewitt$^{\ddag}$}~\affiliation{Rice University, Houston, Texas 77005, USA}
\author{D.~Price$^{\ddag}$}~\affiliation{Indiana University, Bloomington, Indiana 47405, USA}
\author{N.~Prokopenko$^{\ddag}$}~\affiliation{Institute for High Energy Physics, Protvino, Russia}
\author{F.~Prokoshin$^{\dag w}$}~\affiliation{Joint Institute for Nuclear Research, Dubna, Russia}
\author{F.~Ptohos$^{\dag x}$}~\affiliation{Laboratori Nazionali di Frascati, Istituto Nazionale di Fisica Nucleare, I-00044 Frascati, Italy}
\author{G.~Punzi$^{\dag\sharp{c}}$}~\affiliation{Istituto Nazionale di Fisica Nucleare Pisa, $^{\sharp{c}}$University of Pisa, $^{\sharp{d}}$University of Siena and $^{\sharp{e}}$Scuola Normale Superiore, I-56127 Pisa, Italy}
\author{J.~Qian$^{\ddag}$}~\affiliation{University of Michigan, Ann Arbor, Michigan 48109, USA}
\author{A.~Quadt$^{\ddag}$}~\affiliation{II. Physikalisches Institut, Georg-August-Universit\"at G\"ottingen, G\"ottingen, Germany}
\author{B.~Quinn$^{\ddag}$}~\affiliation{University of Mississippi, University, Mississippi 38677, USA}
\author{A.~Rahaman$^{\dag}$}~\affiliation{University of Pittsburgh, Pittsburgh, Pennsylvania 15260, USA}
\author{V.~Ramakrishnan$^{\dag}$}~\affiliation{University of Wisconsin, Madison, Wisconsin 53706, USA}
\author{M.S.~Rangel$^{\ddag}$}~\affiliation{LAFEX, Centro Brasileiro de Pesquisas F\'{i}sicas, Rio de Janeiro, Brazil}
\author{K.~Ranjan$^{\ddag}$}~\affiliation{Delhi University, Delhi, India}
\author{N.~Ranjan$^{\dag}$}~\affiliation{Purdue University, West Lafayette, Indiana 47907, USA}
\author{P.N.~Ratoff$^{\ddag}$}~\affiliation{Lancaster University, Lancaster LA1 4YB, United Kingdom}
\author{I.~Razumov$^{\ddag}$}~\affiliation{Institute for High Energy Physics, Protvino, Russia}
\author{I.~Redondo$^{\dag}$}~\affiliation{Centro de Investigaciones Energeticas Medioambientales y Tecnologicas, E-28040 Madrid, Spain}
\author{P.~Renkel$^{\ddag}$}~\affiliation{Southern Methodist University, Dallas, Texas 75275, USA}
\author{P.~Renton$^{\dag}$}~\affiliation{University of Oxford, Oxford OX1 3RH, United Kingdom}
\author{M.~Rescigno$^{\dag}$}~\affiliation{Istituto Nazionale di Fisica Nucleare, Sezione di Roma 1, $^{\sharp{f}}$Sapienza Universit\`{a} di Roma, I-00185 Roma, Italy}
\author{T.~Riddick$^{\dag}$}~\affiliation{University College London, London WC1E 6BT, United Kingdom}
\author{F.~Rimondi$^{\dag\sharp{a}}$}~\affiliation{Istituto Nazionale di Fisica Nucleare Bologna, $^{\sharp{a}}$University of Bologna, I-40127 Bologna, Italy}
\author{I.~Ripp-Baudot$^{\ddag}$}~\affiliation{IPHC, Universit\'e de Strasbourg, CNRS/IN2P3, Strasbourg, France}
\author{L.~Ristori$^{\dag}$}\affiliation{University of Pittsburgh, Pittsburgh, Pennsylvania 15260, USA}~\affiliation{Fermi National Accelerator Laboratory, Batavia, Illinois 60510, USA}
\author{F.~Rizatdinova$^{\ddag}$}~\affiliation{Oklahoma State University, Stillwater, Oklahoma 74078, USA}
\author{A.~Robson$^{\dag}$}~\affiliation{Glasgow University, Glasgow G12 8QQ, United Kingdom}
\author{T.~Rodrigo$^{\dag}$}~\affiliation{Instituto de Fisica de Cantabria, CSIC-University of Cantabria, 39005 Santander, Spain}
\author{T.~Rodriguez$^{\dag}$}~\affiliation{University of Pennsylvania, Philadelphia, Pennsylvania 19104, USA}
\author{E.~Rogers$^{\dag}$}~\affiliation{University of Illinois, Urbana, Illinois 61801, USA}
\author{S.~Rolli$^{\dag y}$}~\affiliation{Tufts University, Medford, Massachusetts 02155, USA}
\author{M.~Rominsky$^{\ddag}$}~\affiliation{Fermi National Accelerator Laboratory, Batavia, Illinois 60510, USA}
\author{R.~Roser$^{\dag}$}~\affiliation{Fermi National Accelerator Laboratory, Batavia, Illinois 60510, USA}
\author{A.~Ross$^{\ddag}$}~\affiliation{Lancaster University, Lancaster LA1 4YB, United Kingdom}
\author{C.~Royon$^{\ddag}$}~\affiliation{CEA, Irfu, SPP, Saclay, France}
\author{P.~Rubinov$^{\ddag}$}~\affiliation{Fermi National Accelerator Laboratory, Batavia, Illinois 60510, USA}
\author{R.~Ruchti$^{\ddag}$}~\affiliation{University of Notre Dame, Notre Dame, Indiana 46556, USA}
\author{F.~Ruffini$^{\dag\sharp{d}}$}~\affiliation{Istituto Nazionale di Fisica Nucleare Pisa, $^{\sharp{c}}$University of Pisa, $^{\sharp{d}}$University of Siena and $^{\sharp{e}}$Scuola Normale Superiore, I-56127 Pisa, Italy}
\author{A.~Ruiz$^{\dag}$}~\affiliation{Instituto de Fisica de Cantabria, CSIC-University of Cantabria, 39005 Santander, Spain}
\author{J.~Russ$^{\dag}$}~\affiliation{Carnegie Mellon University, Pittsburgh, Pennsylvania 15213, USA}
\author{V.~Rusu$^{\dag}$}~\affiliation{Fermi National Accelerator Laboratory, Batavia, Illinois 60510, USA}
\author{A.~Safonov$^{\dag}$}~\affiliation{Texas A\&M University, College Station, Texas 77843, USA}
\author{G.~Sajot$^{\ddag}$}~\affiliation{LPSC, Universit\'e Joseph Fourier Grenoble 1, CNRS/IN2P3, Institut National Polytechnique de Grenoble, Grenoble, France}
\author{W.K.~Sakumoto$^{\dag}$}~\affiliation{University of Rochester, Rochester, New York 14627, USA}
\author{Y.~Sakurai$^{\dag}$}~\affiliation{Waseda University, Tokyo 169, Japan}
\author{P.~Salcido$^{\ddag}$}~\affiliation{Northern Illinois University, DeKalb, Illinois 60115, USA}
\author{A.~S\'anchez-Hern\'andez$^{\ddag}$}~\affiliation{CINVESTAV, Mexico City, Mexico}
\author{M.P.~Sanders$^{\ddag}$}~\affiliation{Ludwig-Maximilians-Universit\"at M\"unchen, M\"unchen, Germany}
\author{L.~Santi$^{\dag\sharp{g}}$}~\affiliation{Istituto Nazionale di Fisica Nucleare Trieste/Udine, I-34100 Trieste, $^{\sharp{g}}$University of Udine, I-33100 Udine, Italy}
\author{A.S.~Santos$^{\ddag i}$}~\affiliation{LAFEX, Centro Brasileiro de Pesquisas F\'{i}sicas, Rio de Janeiro, Brazil}
\author{K.~Sato$^{\dag}$}~\affiliation{University of Tsukuba, Tsukuba, Ibaraki 305, Japan}
\author{G.~Savage$^{\ddag}$}~\affiliation{Fermi National Accelerator Laboratory, Batavia, Illinois 60510, USA}
\author{V.~Saveliev$^{\dag g}$}~\affiliation{Fermi National Accelerator Laboratory, Batavia, Illinois 60510, USA}
\author{A.~Savoy-Navarro$^{\dag z}$}~\affiliation{Fermi National Accelerator Laboratory, Batavia, Illinois 60510, USA}
\author{L.~Sawyer$^{\ddag}$}~\affiliation{Louisiana Tech University, Ruston, Louisiana 71272, USA}
\author{T.~Scanlon$^{\ddag}$}~\affiliation{Imperial College London, London SW7 2AZ, United Kingdom}
\author{R.D.~Schamberger$^{\ddag}$}~\affiliation{State University of New York, Stony Brook, New York 11794, USA}
\author{Y.~Scheglov$^{\ddag}$}~\affiliation{Petersburg Nuclear Physics Institute, St. Petersburg, Russia}
\author{H.~Schellman$^{\ddag}$}~\affiliation{Northwestern University, Evanston, Illinois 60208, USA}
\author{P.~Schlabach$^{\dag}$}~\affiliation{Fermi National Accelerator Laboratory, Batavia, Illinois 60510, USA}
\author{S.~Schlobohm$^{\ddag}$}~\affiliation{University of Washington, Seattle, Washington 98195, USA}
\author{A.~Schmidt$^{\dag}$}~\affiliation{Institut f\"{u}r Experimentelle Kernphysik, Karlsruhe Institute of Technology, D-76131 Karlsruhe, Germany}
\author{E.E.~Schmidt$^{\dag}$}~\affiliation{Fermi National Accelerator Laboratory, Batavia, Illinois 60510, USA}
\author{C.~Schwanenberger$^{\ddag}$}~\affiliation{The University of Manchester, Manchester M13 9PL, United Kingdom}
\author{T.~Schwarz$^{\dag}$}~\affiliation{Fermi National Accelerator Laboratory, Batavia, Illinois 60510, USA}
\author{R.~Schwienhorst$^{\ddag}$}~\affiliation{Michigan State University, East Lansing, Michigan 48824, USA}
\author{L.~Scodellaro$^{\dag}$}~\affiliation{Instituto de Fisica de Cantabria, CSIC-University of Cantabria, 39005 Santander, Spain}
\author{A.~Scribano$^{\dag\sharp{d}}$}~\affiliation{Istituto Nazionale di Fisica Nucleare Pisa, $^{\sharp{c}}$University of Pisa, $^{\sharp{d}}$University of Siena and $^{\sharp{e}}$Scuola Normale Superiore, I-56127 Pisa, Italy}
\author{F.~Scuri$^{\dag}$}~\affiliation{Istituto Nazionale di Fisica Nucleare Pisa, $^{\sharp{c}}$University of Pisa, $^{\sharp{d}}$University of Siena and $^{\sharp{e}}$Scuola Normale Superiore, I-56127 Pisa, Italy}
\author{S.~Seidel$^{\dag}$}~\affiliation{University of New Mexico, Albuquerque, New Mexico 87131, USA}
\author{Y.~Seiya$^{\dag}$}~\affiliation{Osaka City University, Osaka 588, Japan}
\author{J.~Sekaric$^{\ddag}$}~\affiliation{University of Kansas, Lawrence, Kansas 66045, USA}
\author{A.~Semenov$^{\dag}$}~\affiliation{Joint Institute for Nuclear Research, Dubna, Russia}
\author{H.~Severini$^{\ddag}$}~\affiliation{University of Oklahoma, Norman, Oklahoma 73019, USA}
\author{F.~Sforza$^{\dag\sharp{d}}$}~\affiliation{Istituto Nazionale di Fisica Nucleare Pisa, $^{\sharp{c}}$University of Pisa, $^{\sharp{d}}$University of Siena and $^{\sharp{e}}$Scuola Normale Superiore, I-56127 Pisa, Italy}
\author{E.~Shabalina$^{\ddag}$}~\affiliation{II. Physikalisches Institut, Georg-August-Universit\"at G\"ottingen, G\"ottingen, Germany}
\author{S.Z.~Shalhout$^{\dag}$}~\affiliation{University of California, Davis, Davis, California 95616, USA}
\author{V.~Shary$^{\ddag}$}~\affiliation{CEA, Irfu, SPP, Saclay, France}
\author{S.~Shaw$^{\ddag}$}~\affiliation{Michigan State University, East Lansing, Michigan 48824, USA}
\author{A.A.~Shchukin$^{\ddag}$}~\affiliation{Institute for High Energy Physics, Protvino, Russia}
\author{T.~Shears$^{\dag}$}~\affiliation{University of Liverpool, Liverpool L69 7ZE, United Kingdom}
\author{P.F.~Shepard$^{\dag}$}~\affiliation{University of Pittsburgh, Pittsburgh, Pennsylvania 15260, USA}
\author{M.~Shimojima$^{\dag aa}$}~\affiliation{University of Tsukuba, Tsukuba, Ibaraki 305, Japan}
\author{R.K.~Shivpuri$^{\ddag}$}~\affiliation{Delhi University, Delhi, India}
\author{M.~Shochet$^{\dag}$}~\affiliation{Enrico Fermi Institute, University of Chicago, Chicago, Illinois 60637, USA}
\author{I.~Shreyber-Tecker$^{\dag}$}~\affiliation{Institution for Theoretical and Experimental Physics, ITEP, Moscow 117259, Russia}
\author{V.~Simak$^{\ddag}$}~\affiliation{Czech Technical University in Prague, Prague, Czech Republic}
\author{A.~Simonenko$^{\dag}$}~\affiliation{Joint Institute for Nuclear Research, Dubna, Russia}
\author{P.~Sinervo$^{\dag}$}~\affiliation{Institute of Particle Physics: McGill University, Montr\'{e}al, Qu\'{e}bec, Canada H3A~2T8; Simon Fraser University, Burnaby, British Columbia, Canada V5A~1S6; University of Toronto, Toronto, Ontario, Canada M5S~1A7; and TRIUMF, Vancouver, British Columbia, Canada V6T~2A3}
\author{P.~Skubic$^{\ddag}$}~\affiliation{University of Oklahoma, Norman, Oklahoma 73019, USA}
\author{P.~Slattery$^{\ddag}$}~\affiliation{University of Rochester, Rochester, New York 14627, USA}
\author{K.~Sliwa$^{\dag}$}~\affiliation{Tufts University, Medford, Massachusetts 02155, USA}
\author{D.~Smirnov$^{\ddag}$}~\affiliation{University of Notre Dame, Notre Dame, Indiana 46556, USA}
\author{J.R.~Smith$^{\dag}$}~\affiliation{University of California, Davis, Davis, California 95616, USA}
\author{K.J.~Smith$^{\ddag}$}~\affiliation{State University of New York, Buffalo, New York 14260, USA}
\author{F.D.~Snider$^{\dag}$}~\affiliation{Fermi National Accelerator Laboratory, Batavia, Illinois 60510, USA}
\author{G.R.~Snow$^{\ddag}$}~\affiliation{University of Nebraska, Lincoln, Nebraska 68588, USA}
\author{J.~Snow$^{\ddag}$}~\affiliation{Langston University, Langston, Oklahoma 73050, USA}
\author{S.~Snyder$^{\ddag}$}~\affiliation{Brookhaven National Laboratory, Upton, New York 11973, USA}
\author{A.~Soha$^{\dag}$}~\affiliation{Fermi National Accelerator Laboratory, Batavia, Illinois 60510, USA}
\author{S.~S{\"o}ldner-Rembold$^{\ddag}$}~\affiliation{The University of Manchester, Manchester M13 9PL, United Kingdom}
\author{H.~Song$^{\dag}$}~\affiliation{University of Pittsburgh, Pittsburgh, Pennsylvania 15260, USA}
\author{L.~Sonnenschein$^{\ddag}$}~\affiliation{III. Physikalisches Institut A, RWTH Aachen University, Aachen, Germany}
\author{V.~Sorin$^{\dag}$}~\affiliation{Institut de Fisica d'Altes Energies, ICREA, Universitat Autonoma de Barcelona, E-08193, Bellaterra (Barcelona), Spain}
\author{K.~Soustruznik$^{\ddag}$}~\affiliation{Charles University, Faculty of Mathematics and Physics, Center for Particle Physics, Prague, Czech Republic}
\author{P.~Squillacioti$^{\dag\sharp{d}}$}~\affiliation{Istituto Nazionale di Fisica Nucleare Pisa, $^{\sharp{c}}$University of Pisa, $^{\sharp{d}}$University of Siena and $^{\sharp{e}}$Scuola Normale Superiore, I-56127 Pisa, Italy}
\author{R.~St.~Denis$^{\dag}$}~\affiliation{Glasgow University, Glasgow G12 8QQ, United Kingdom}
\author{M.~Stancari$^{\dag}$}~\affiliation{Fermi National Accelerator Laboratory, Batavia, Illinois 60510, USA}
\author{J.~Stark$^{\ddag}$}~\affiliation{LPSC, Universit\'e Joseph Fourier Grenoble 1, CNRS/IN2P3, Institut National Polytechnique de Grenoble, Grenoble, France}
\author{O.~Stelzer-Chilton$^{\dag}$}~\affiliation{Institute of Particle Physics: McGill University, Montr\'{e}al, Qu\'{e}bec, Canada H3A~2T8; Simon Fraser University, Burnaby, British Columbia, Canada V5A~1S6; University of Toronto, Toronto, Ontario, Canada M5S~1A7; and TRIUMF, Vancouver, British Columbia, Canada V6T~2A3}
\author{B.~Stelzer$^{\dag}$}~\affiliation{Institute of Particle Physics: McGill University, Montr\'{e}al, Qu\'{e}bec, Canada H3A~2T8; Simon Fraser University, Burnaby, British Columbia, Canada V5A~1S6; University of Toronto, Toronto, Ontario, Canada M5S~1A7; and TRIUMF, Vancouver, British Columbia, Canada V6T~2A3}
\author{D.~Stentz$^{\dag b}$}~\affiliation{Fermi National Accelerator Laboratory, Batavia, Illinois 60510, USA}
\author{D.A.~Stoyanova$^{\ddag}$}~\affiliation{Institute for High Energy Physics, Protvino, Russia}
\author{M.~Strauss$^{\ddag}$}~\affiliation{University of Oklahoma, Norman, Oklahoma 73019, USA}
\author{J.~Strologas$^{\dag}$}~\affiliation{University of New Mexico, Albuquerque, New Mexico 87131, USA}
\author{G.L.~Strycker$^{\dag}$}~\affiliation{University of Michigan, Ann Arbor, Michigan 48109, USA}
\author{Y.~Sudo$^{\dag}$}~\affiliation{University of Tsukuba, Tsukuba, Ibaraki 305, Japan}
\author{A.~Sukhanov$^{\dag}$}~\affiliation{Fermi National Accelerator Laboratory, Batavia, Illinois 60510, USA}
\author{I.~Suslov$^{\dag}$}~\affiliation{Joint Institute for Nuclear Research, Dubna, Russia}
\author{L.~Suter$^{\ddag}$}~\affiliation{The University of Manchester, Manchester M13 9PL, United Kingdom}
\author{P.~Svoisky$^{\ddag}$}~\affiliation{University of Oklahoma, Norman, Oklahoma 73019, USA}
\author{M.~Takahashi$^{\ddag}$}~\affiliation{The University of Manchester, Manchester M13 9PL, United Kingdom}
\author{K.~Takemasa$^{\dag}$}~\affiliation{University of Tsukuba, Tsukuba, Ibaraki 305, Japan}
\author{Y.~Takeuchi$^{\dag}$}~\affiliation{University of Tsukuba, Tsukuba, Ibaraki 305, Japan}
\author{J.~Tang$^{\dag}$}~\affiliation{Enrico Fermi Institute, University of Chicago, Chicago, Illinois 60637, USA}
\author{M.~Tecchio$^{\dag}$}~\affiliation{University of Michigan, Ann Arbor, Michigan 48109, USA}
\author{P.K.~Teng$^{\dag}$}~\affiliation{Institute of Physics, Academia Sinica, Taipei, Taiwan 11529, Republic of China}
\author{J.~Thom$^{\dag j}$}~\affiliation{Fermi National Accelerator Laboratory, Batavia, Illinois 60510, USA}
\author{J.~Thome$^{\dag}$}~\affiliation{Carnegie Mellon University, Pittsburgh, Pennsylvania 15213, USA}
\author{G.A.~Thompson$^{\dag}$}~\affiliation{University of Illinois, Urbana, Illinois 61801, USA}
\author{E.~Thomson$^{\dag}$}~\affiliation{University of Pennsylvania, Philadelphia, Pennsylvania 19104, USA}
\author{M.~Titov$^{\ddag}$}~\affiliation{CEA, Irfu, SPP, Saclay, France}
\author{D.~Toback$^{\dag}$}~\affiliation{Texas A\&M University, College Station, Texas 77843, USA}
\author{S.~Tokar$^{\dag}$}~\affiliation{Comenius University, 842 48 Bratislava, Slovakia; Institute of Experimental Physics, 040 01 Kosice, Slovakia}
\author{V.V.~Tokmenin$^{\ddag}$}~\affiliation{Joint Institute for Nuclear Research, Dubna, Russia}
\author{K.~Tollefson$^{\dag}$}~\affiliation{Michigan State University, East Lansing, Michigan 48824, USA}
\author{T.~Tomura$^{\dag}$}~\affiliation{University of Tsukuba, Tsukuba, Ibaraki 305, Japan}
\author{D.~Tonelli$^{\dag}$}~\affiliation{Fermi National Accelerator Laboratory, Batavia, Illinois 60510, USA}
\author{S.~Torre$^{\dag}$}~\affiliation{Laboratori Nazionali di Frascati, Istituto Nazionale di Fisica Nucleare, I-00044 Frascati, Italy}
\author{D.~Torretta$^{\dag}$}~\affiliation{Fermi National Accelerator Laboratory, Batavia, Illinois 60510, USA}
\author{P.~Totaro$^{\dag}$}~\affiliation{Istituto Nazionale di Fisica Nucleare, Sezione di Padova-Trento, $^{\sharp{b}}$University of Padova, I-35131 Padova, Italy}
\author{M.~Trovato$^{\dag\sharp{e}}$}~\affiliation{Istituto Nazionale di Fisica Nucleare Pisa, $^{\sharp{c}}$University of Pisa, $^{\sharp{d}}$University of Siena and $^{\sharp{e}}$Scuola Normale Superiore, I-56127 Pisa, Italy}
\author{Y.-T.~Tsai$^{\ddag}$}~\affiliation{University of Rochester, Rochester, New York 14627, USA}
\author{K.~Tschann-Grimm$^{\ddag}$}~\affiliation{State University of New York, Stony Brook, New York 11794, USA}
\author{D.~Tsybychev$^{\ddag}$}~\affiliation{State University of New York, Stony Brook, New York 11794, USA}
\author{B.~Tuchming$^{\ddag}$}~\affiliation{CEA, Irfu, SPP, Saclay, France}
\author{C.~Tully$^{\ddag}$}~\affiliation{Princeton University, Princeton, New Jersey 08544, USA}
\author{F.~Ukegawa$^{\dag}$}~\affiliation{University of Tsukuba, Tsukuba, Ibaraki 305, Japan}
\author{S.~Uozumi$^{\dag}$}~\affiliation{Center for High Energy Physics: Kyungpook National University, Daegu 702-701, Korea; Seoul National University, Seoul 151-742, Korea; Sungkyunkwan University, Suwon 440-746, Korea; Korea Institute of Science and Technology Information, Daejeon 305-806, Korea; Chonnam National University, Gwangju 500-757, Korea; Chonbuk National University, Jeonju 561-756, Korea}
\author{L.~Uvarov$^{\ddag}$}~\affiliation{Petersburg Nuclear Physics Institute, St. Petersburg, Russia}
\author{S.~Uvarov$^{\ddag}$}~\affiliation{Petersburg Nuclear Physics Institute, St. Petersburg, Russia}
\author{S.~Uzunyan$^{\ddag}$}~\affiliation{Northern Illinois University, DeKalb, Illinois 60115, USA}
\author{R.~Van~Kooten$^{\ddag}$}~\affiliation{Indiana University, Bloomington, Indiana 47405, USA}
\author{W.M.~van~Leeuwen$^{\ddag}$}~\affiliation{Nikhef, Science Park, Amsterdam, the Netherlands}
\author{N.~Varelas$^{\ddag}$}~\affiliation{University of Illinois at Chicago, Chicago, Illinois 60607, USA}
\author{A.~Varganov$^{\dag}$}~\affiliation{University of Michigan, Ann Arbor, Michigan 48109, USA}
\author{E.W.~Varnes$^{\ddag}$}~\affiliation{University of Arizona, Tucson, Arizona 85721, USA}
\author{I.A.~Vasilyev$^{\ddag}$}~\affiliation{Institute for High Energy Physics, Protvino, Russia}
\author{F.~V\'{a}zquez$^{\dag c}$}~\affiliation{University of Florida, Gainesville, Florida 32611, USA}
\author{G.~Velev$^{\dag}$}~\affiliation{Fermi National Accelerator Laboratory, Batavia, Illinois 60510, USA}
\author{C.~Vellidis$^{\dag}$}~\affiliation{Fermi National Accelerator Laboratory, Batavia, Illinois 60510, USA}
\author{P.~Verdier$^{\ddag}$}~\affiliation{IPNL, Universit\'e Lyon 1, CNRS/IN2P3, Villeurbanne, France and Universit\'e de Lyon, Lyon, France}
\author{A.Y.~Verkheev$^{\ddag}$}~\affiliation{Joint Institute for Nuclear Research, Dubna, Russia}
\author{L.S.~Vertogradov$^{\ddag}$}~\affiliation{Joint Institute for Nuclear Research, Dubna, Russia}
\author{M.~Verzocchi$^{\ddag}$}~\affiliation{Fermi National Accelerator Laboratory, Batavia, Illinois 60510, USA}
\author{M.~Vesterinen$^{\ddag}$}~\affiliation{The University of Manchester, Manchester M13 9PL, United Kingdom}
\author{M.~Vidal$^{\dag}$}~\affiliation{Purdue University, West Lafayette, Indiana 47907, USA}
\author{I.~Vila$^{\dag}$}~\affiliation{Instituto de Fisica de Cantabria, CSIC-University of Cantabria, 39005 Santander, Spain}
\author{D.~Vilanova$^{\ddag}$}~\affiliation{CEA, Irfu, SPP, Saclay, France}
\author{R.~Vilar$^{\dag}$}~\affiliation{Instituto de Fisica de Cantabria, CSIC-University of Cantabria, 39005 Santander, Spain}
\author{J.~Viz\'{a}n$^{\dag}$}~\affiliation{Instituto de Fisica de Cantabria, CSIC-University of Cantabria, 39005 Santander, Spain}
\author{M.~Vogel$^{\dag}$}~\affiliation{University of New Mexico, Albuquerque, New Mexico 87131, USA}
\author{P.~Vokac$^{\ddag}$}~\affiliation{Czech Technical University in Prague, Prague, Czech Republic}
\author{G.~Volpi$^{\dag}$}~\affiliation{Laboratori Nazionali di Frascati, Istituto Nazionale di Fisica Nucleare, I-00044 Frascati, Italy}
\author{P.~Wagner$^{\dag}$}~\affiliation{University of Pennsylvania, Philadelphia, Pennsylvania 19104, USA}
\author{R.L.~Wagner$^{\dag}$}~\affiliation{Fermi National Accelerator Laboratory, Batavia, Illinois 60510, USA}
\author{H.D.~Wahl$^{\ddag}$}~\affiliation{Florida State University, Tallahassee, Florida 32306, USA}
\author{T.~Wakisaka$^{\dag}$}~\affiliation{Osaka City University, Osaka 588, Japan}
\author{R.~Wallny$^{\dag}$}~\affiliation{University of California, Los Angeles, Los Angeles, California 90024, USA}
\author{S.M.~Wang$^{\dag}$}~\affiliation{Institute of Physics, Academia Sinica, Taipei, Taiwan 11529, Republic of China}
\author{M.H.L.S.~Wang$^{\ddag}$}~\affiliation{Fermi National Accelerator Laboratory, Batavia, Illinois 60510, USA}
\author{A.~Warburton$^{\dag}$}~\affiliation{Institute of Particle Physics: McGill University, Montr\'{e}al, Qu\'{e}bec, Canada H3A~2T8; Simon Fraser University, Burnaby, British Columbia, Canada V5A~1S6; University of Toronto, Toronto, Ontario, Canada M5S~1A7; and TRIUMF, Vancouver, British Columbia, Canada V6T~2A3}
\author{J.~Warchol$^{\ddag}$}~\affiliation{University of Notre Dame, Notre Dame, Indiana 46556, USA}
\author{D.~Waters$^{\dag}$}~\affiliation{University College London, London WC1E 6BT, United Kingdom}
\author{G.~Watts$^{\ddag}$}~\affiliation{University of Washington, Seattle, Washington 98195, USA}
\author{M.~Wayne$^{\ddag}$}~\affiliation{University of Notre Dame, Notre Dame, Indiana 46556, USA}
\author{J.~Weichert$^{\ddag}$}~\affiliation{Institut f\"ur Physik, Universit\"at Mainz, Mainz, Germany}
\author{L.~Welty-Rieger$^{\ddag}$}~\affiliation{Northwestern University, Evanston, Illinois 60208, USA}
\author{W.C.~Wester~III$^{\dag}$}~\affiliation{Fermi National Accelerator Laboratory, Batavia, Illinois 60510, USA}
\author{A.~White$^{\ddag}$}~\affiliation{University of Texas, Arlington, Texas 76019, USA}
\author{D.~Whiteson$^{\dag bb}$}~\affiliation{University of Pennsylvania, Philadelphia, Pennsylvania 19104, USA}
\author{F.~Wick$^{\dag}$}~\affiliation{Institut f\"{u}r Experimentelle Kernphysik, Karlsruhe Institute of Technology, D-76131 Karlsruhe, Germany}
\author{D.~Wicke$^{\ddag}$}~\affiliation{Fachbereich Physik, Bergische Universit\"at Wuppertal, Wuppertal, Germany}
\author{A.B.~Wicklund$^{\dag}$}~\affiliation{Argonne National Laboratory, Argonne, Illinois 60439, USA}
\author{E.~Wicklund$^{\dag}$}~\affiliation{Fermi National Accelerator Laboratory, Batavia, Illinois 60510, USA}
\author{S.~Wilbur$^{\dag}$}~\affiliation{Enrico Fermi Institute, University of Chicago, Chicago, Illinois 60637, USA}
\author{H.H.~Williams$^{\dag}$}~\affiliation{University of Pennsylvania, Philadelphia, Pennsylvania 19104, USA}
\author{M.R.J.~Williams$^{\ddag}$}~\affiliation{Lancaster University, Lancaster LA1 4YB, United Kingdom}
\author{G.W.~Wilson$^{\ddag}$}~\affiliation{University of Kansas, Lawrence, Kansas 66045, USA}
\author{J.S.~Wilson$^{\dag}$}~\affiliation{The Ohio State University, Columbus, Ohio 43210, USA}
\author{P.~Wilson$^{\dag}$}~\affiliation{Fermi National Accelerator Laboratory, Batavia, Illinois 60510, USA}
\author{B.L.~Winer$^{\dag}$}~\affiliation{The Ohio State University, Columbus, Ohio 43210, USA}
\author{P.~Wittich$^{\dag j}$}~\affiliation{Fermi National Accelerator Laboratory, Batavia, Illinois 60510, USA}
\author{M.~Wobisch$^{\ddag}$}~\affiliation{Louisiana Tech University, Ruston, Louisiana 71272, USA}
\author{S.~Wolbers$^{\dag}$}~\affiliation{Fermi National Accelerator Laboratory, Batavia, Illinois 60510, USA}
\author{H.~Wolfe$^{\dag}$}~\affiliation{The Ohio State University, Columbus, Ohio 43210, USA}
\author{D.R.~Wood$^{\ddag}$}~\affiliation{Northeastern University, Boston, Massachusetts 02115, USA}
\author{T.~Wright$^{\dag}$}~\affiliation{University of Michigan, Ann Arbor, Michigan 48109, USA}
\author{X.~Wu$^{\dag}$}~\affiliation{University of Geneva, CH-1211 Geneva 4, Switzerland}
\author{Z.~Wu$^{\dag}$}~\affiliation{Baylor University, Waco, Texas 76798, USA}
\author{T.R.~Wyatt$^{\ddag}$}~\affiliation{The University of Manchester, Manchester M13 9PL, United Kingdom}
\author{Y.~Xie$^{\ddag}$}~\affiliation{Fermi National Accelerator Laboratory, Batavia, Illinois 60510, USA}
\author{R.~Yamada$^{\ddag}$}~\affiliation{Fermi National Accelerator Laboratory, Batavia, Illinois 60510, USA}
\author{K.~Yamamoto$^{\dag}$}~\affiliation{Osaka City University, Osaka 588, Japan}
\author{D.~Yamato$^{\dag}$}~\affiliation{Osaka City University, Osaka 588, Japan}
\author{S.~Yang$^{\ddag}$}~\affiliation{University of Science and Technology of China, Hefei, People's Republic of China}
\author{T.~Yang$^{\dag}$}~\affiliation{Fermi National Accelerator Laboratory, Batavia, Illinois 60510, USA}
\author{U.K.~Yang$^{\dag cc}$}~\affiliation{Enrico Fermi Institute, University of Chicago, Chicago, Illinois 60637, USA}
\author{W.-C.~Yang$^{\ddag}$}~\affiliation{The University of Manchester, Manchester M13 9PL, United Kingdom}
\author{Y.C.~Yang$^{\dag}$}~\affiliation{Center for High Energy Physics: Kyungpook National University, Daegu 702-701, Korea; Seoul National University, Seoul 151-742, Korea; Sungkyunkwan University, Suwon 440-746, Korea; Korea Institute of Science and Technology Information, Daejeon 305-806, Korea; Chonnam National University, Gwangju 500-757, Korea; Chonbuk National University, Jeonju 561-756, Korea}
\author{W.-M.~Yao$^{\dag}$}~\affiliation{Ernest Orlando Lawrence Berkeley National Laboratory, Berkeley, California 94720, USA}
\author{T.~Yasuda$^{\ddag}$}~\affiliation{Fermi National Accelerator Laboratory, Batavia, Illinois 60510, USA}
\author{Y.A.~Yatsunenko$^{\ddag}$}~\affiliation{Joint Institute for Nuclear Research, Dubna, Russia}
\author{W.~Ye$^{\ddag}$}~\affiliation{State University of New York, Stony Brook, New York 11794, USA}
\author{Z.~Ye$^{\ddag}$}~\affiliation{Fermi National Accelerator Laboratory, Batavia, Illinois 60510, USA}
\author{G.P.~Yeh$^{\dag}$}~\affiliation{Fermi National Accelerator Laboratory, Batavia, Illinois 60510, USA}
\author{K.~Yi$^{\dag u}$}~\affiliation{Fermi National Accelerator Laboratory, Batavia, Illinois 60510, USA}
\author{H.~Yin$^{\ddag}$}~\affiliation{Fermi National Accelerator Laboratory, Batavia, Illinois 60510, USA}
\author{K.~Yip$^{\ddag}$}~\affiliation{Brookhaven National Laboratory, Upton, New York 11973, USA}
\author{J.~Yoh$^{\dag}$}~\affiliation{Fermi National Accelerator Laboratory, Batavia, Illinois 60510, USA}
\author{K.~Yorita$^{\dag}$}~\affiliation{Waseda University, Tokyo 169, Japan}
\author{T.~Yoshida$^{\dag dd}$}~\affiliation{Osaka City University, Osaka 588, Japan}
\author{S.W.~Youn$^{\ddag}$}~\affiliation{Fermi National Accelerator Laboratory, Batavia, Illinois 60510, USA}
\author{G.B.~Yu$^{\dag}$}~\affiliation{Duke University, Durham, North Carolina 27708, USA}
\author{I.~Yu$^{\dag}$}~\affiliation{Center for High Energy Physics: Kyungpook National University, Daegu 702-701, Korea; Seoul National University, Seoul 151-742, Korea; Sungkyunkwan University, Suwon 440-746, Korea; Korea Institute of Science and Technology Information, Daejeon 305-806, Korea; Chonnam National University, Gwangju 500-757, Korea; Chonbuk National University, Jeonju 561-756, Korea}
\author{J.M.~Yu$^{\ddag}$}~\affiliation{University of Michigan, Ann Arbor, Michigan 48109, USA}
\author{S.S.~Yu$^{\dag}$}~\affiliation{Fermi National Accelerator Laboratory, Batavia, Illinois 60510, USA}
\author{J.C.~Yun$^{\dag}$}~\affiliation{Fermi National Accelerator Laboratory, Batavia, Illinois 60510, USA}
\author{A.~Zanetti$^{\dag}$}~\affiliation{Istituto Nazionale di Fisica Nucleare Trieste/Udine, I-34100 Trieste, $^{\sharp{g}}$University of Udine, I-33100 Udine, Italy}
\author{Y.~Zeng$^{\dag}$}~\affiliation{Duke University, Durham, North Carolina 27708, USA}
\author{J.~Zennamo$^{\ddag}$}~\affiliation{State University of New York, Buffalo, New York 14260, USA}
\author{T.~Zhao$^{\ddag}$}~\affiliation{University of Washington, Seattle, Washington 98195, USA}
\author{T.G.~Zhao$^{\ddag}$}~\affiliation{The University of Manchester, Manchester M13 9PL, United Kingdom}
\author{B.~Zhou$^{\ddag}$}~\affiliation{University of Michigan, Ann Arbor, Michigan 48109, USA}
\author{C.~Zhou$^{\dag}$}~\affiliation{Duke University, Durham, North Carolina 27708, USA}
\author{J.~Zhu$^{\ddag}$}~\affiliation{University of Michigan, Ann Arbor, Michigan 48109, USA}
\author{M.~Zielinski$^{\ddag}$}~\affiliation{University of Rochester, Rochester, New York 14627, USA}
\author{D.~Zieminska$^{\ddag}$}~\affiliation{Indiana University, Bloomington, Indiana 47405, USA}
\author{L.~Zivkovic$^{\ddag}$}~\affiliation{Brown University, Providence, Rhode Island 02912, USA}
\author{S.~Zucchelli$^{\dag\sharp{a}}$}~\affiliation{Istituto Nazionale di Fisica Nucleare Bologna, $^{\sharp{a}}$University of Bologna, I-40127 Bologna, Italy}
\collaboration{CDF\footnote{
With CDF visitors from
$^{{\dag}a}$Universidad de Oviedo, E-33007 Oviedo, Spain,
$^{{\dag}b}$Northwestern University, Evanston, IL 60208, USA,
$^{{\dag}c}$Universidad Iberoamericana, Mexico D.F., Mexico,
$^{{\dag}d}$ETH, 8092 Zurich, Switzerland,
$^{{\dag}e}$CERN, CH-1211 Geneva, Switzerland,
$^{{\dag}f}$Queen Mary, University of London, London, E1 4NS, United Kingdom,
$^{{\dag}g}$National Research Nuclear University, Moscow, Russia,
$^{{\dag}h}$Yarmouk University, Irbid 211-63, Jordan,
$^{{\dag}i}$Muons, Inc., Batavia, IL 60510, USA,
$^{{\dag}j}$Cornell University, Ithaca, NY 14853, USA,
$^{{\dag}k}$Kansas State University, Manhattan, KS 66506, USA,
$^{{\dag}l}$Kinki University, Higashi-Osaka City, Japan 577-8502,
$^{{\dag}m}$University of CA Santa Barbara, Santa Barbara, CA 93106, USA,
$^{{\dag}n}$University of Notre Dame, Notre Dame, IN 46556, USA,
$^{{\dag}o}$Ewha Womans University, Seoul, 120-750, Korea,
$^{{\dag}p}$Texas Tech University, Lubbock, TX 79609, USA,
$^{{\dag}q}$University of Melbourne, Victoria 3010, Australia,
$^{{\dag}r}$Institute of Physics, Academy of Sciences of the Czech Republic, Czech Republic,
$^{{\dag}s}$Istituto Nazionale di Fisica Nucleare, Sezione di Cagliari, 09042 Monserrato (Cagliari), Italy,
$^{{\dag}t}$University College Dublin, Dublin 4, Ireland,
$^{{\dag}u}$University of Iowa, Iowa City, IA 52242, USA,
$^{{\dag}v}$University of CA Santa Cruz, Santa Cruz, CA 95064, USA,
$^{{\dag}w}$Universidad Tecnica Federico Santa Maria, 110v Valparaiso, Chile,
$^{{\dag}x}$University of Cyprus, Nicosia CY-1678, Cyprus,
$^{{\dag}y}$Office of Science, U.S. Department of Energy, Washington, DC 20585, USA,
$^{{\dag}z}$CNRS-IN2P3, Paris, F-75205 France,
$^{{\dag}aa}$Nagasaki Institute of Applied Science, Nagasaki, Japan,
$^{{\dag}bb}$University of CA Irvine, Irvine, CA 92697, USA,
$^{{\dag}cc}$University of Manchester, Manchester M13 9PL, United Kingdom,
and 
$^{{\dag}dd}$University of Fukui, Fukui City, Fukui Prefecture, Japan 910-0017.
} and D0\footnote{
and D0 visitors from
$^{{\ddag}a}$Augustana College, Sioux Falls, SD, USA,
$^{{\ddag}b}$The University of Liverpool, Liverpool, UK,
$^{{\ddag}c}$UPIITA-IPN, Mexico City, Mexico,
$^{{\ddag}d}$DESY, Hamburg, Germany,
$^{{\ddag}e}$University College London, London, UK,
$^{{\ddag}f}$Centro de Investigacion en Computacion - IPN, Mexico City, Mexico,
$^{{\ddag}g}$SLAC, Menlo Park, CA, USA,
$^{{\ddag}h}$ECFM, Universidad Autonoma de Sinaloa, Culiac\'an, Mexico
and
$^{{\ddag}i}$Universidade Estadual Paulista, S\~ao Paulo, Brazil.
} Collaborations}
\noaffiliation

\date{\today}

\begin{abstract}
The combination of searches performed by the CDF and D0 collaborations at the Fermilab Tevatron Collider for neutral Higgs bosons produced in association with $b$ quarks is reported. The data, corresponding to 2.6~\ifb\ of integrated luminosity at CDF and 5.2~\ifb\ at D0, have been collected in final states containing  three or more $b$ jets. Upper limits are set on the cross section multiplied by the branching ratio varying between 44~pb and 0.7~pb in the Higgs boson mass range 90 to 300~GeV, assuming production of a narrow scalar boson. Significant enhancements to the production of Higgs bosons can be found in theories beyond the standard model, for example in supersymmetry. The results are interpreted as upper limits in the parameter space of the minimal supersymmetric standard model in a benchmark scenario favoring this decay mode. 
\end{abstract}

\maketitle 

In the standard model (SM), electroweak symmetry breaking is achieved through the introduction of a single scalar doublet field and the existence of a single neutral scalar boson~\cite{higgs1,higgs2,higgs3,higgs4,higgs5,higgs6} is predicted. However, extensions of the SM exist with richer structure. Introducing a second Higgs doublet, such as in Type II 2-Higgs Doublet Models (2HDM)~\cite{2HDM}, leads to multiple scalar bosons and can give scenarios with enhanced couplings to down-type fermions. Supersymmetry is a plausible extension to the SM that introduces an additional symmetry between fermions and bosons. The two Higgs boson doublets in the minimal supersymmetric standard model (MSSM)~\cite{mssm1,mssm2} lead to five physical Higgs bosons: three neutral (collectively denoted as $\phi$): $h$, $H$, and $A$; and two charged: $H^{+}$ and $H^{-}$. At leading order the MSSM is a Type II 2HDM model, and two parameters are sufficient to describe the Higgs sector. These are conventionally chosen as the ratio of the two Higgs doublet vacuum expectation values, $\tan\beta$, and, $M_A$, the mass of the pseudoscalar boson, $A$. The couplings to the down-type fermions are enhanced by a factor of \tanb\ relative to those in the SM. Thus, the main decay mode is  $\phi\to \b\bbar$, with branching fractions near 90\% at leading order (the remainder being mostly $\phi\to \tau^+\tau^-$). While searches for $\tau$-lepton-pair final states at hadron colliders are relatively insensitive to higher-order radiative corrections, due to cancellations between the production and decay processes, this is not the case for decays to $\b\bbar$. Thus, information from decays to $b\bar{b}$ together with stringent limits from decays to  $\tau^+\tau^-$ can constrain higher-order effects and yield additional information about electroweak symmetry breaking, supersymmetry or other new physics beyond the SM with similar final states such as pair production of color octet scalars~\cite{octet1,octet2,octet3}.

Since an inclusive search for $\phi \to \b\bbar$ is difficult due to large multijet backgrounds, these searches rely on the case where the $\phi$ boson is produced in association with one or more $b$ quarks. This final state with at least three $b$ quarks represents a powerful search channel, with the third $b$ jet providing additional suppression of the large multijet background at a hadron collider. 

MSSM Higgs boson production has been studied at the CERN LEP $e^+e^-$~collider excluding $M_{h,A}<93$~GeV for all $\tan\beta$ values~\cite{cite:LEP_exclu}. The CDF~\cite{cite:CDF_exclu1,cite:CDF_exclu2,cite:CDFbbb} and D0~\cite{cite:D0_exclu, hbbrun2a, cite:D0bbb, tautau1, tautau2, tautau3, tautau4, tautau5, tautau6, d0mssm} collaborations extended such searches to higher masses and for large $\tan\beta$. The most stringent upper limits on \tanb in searches for  neutral Higgs production for masses above the LEP bounds come from searches in final states with $\tau$ leptons pair produced at the Large Hadron Collider~\cite{cite:CMStautau1,cite:ATLAStautau,cite:CMStautau2}. 

The results reported in this Article make use of $p\bar{p}$ collisions at a center of mass energy of 1.96 TeV. The data, corresponding to integrated luminosities of 2.6~\ifb\  and 5.2~\ifb\ , were collected during Run II at the Fermilab Tevatron Collider by the CDF and D0 collaborations, respectively. The combination of searches for neutral Higgs bosons in final states with three or more $b$ jets~\cite{cite:D0bbb,cite:CDFbbb} is presented. 

The CDF and D0 detectors are described in Ref.~\cite{cdfdet,run2det1,run2det2,run2det3}. A brief outline of the reconstruction of the final states and event selection used in these searches follows. A cone algorithm~\cite{cone} with a radius of ${\cal R}=\sqrt{(\Delta y)^2 +(\Delta\varphi)^2}=0.7$~(CDF) and 0.5~(D0), where $y$ is the rapidity and $\varphi$ the azimuthal angle, is used to reconstruct jets from energy deposition in the calorimeters. Identification of jets arising from $b$-quark fragmentation ($b$-tagging) uses an algorithm based on reconstructing secondary vertices of charged particles displaced from the $p\bar{p}$ interaction vertex at CDF~\cite{cdfbtag} and a neural network combining lifetime information from secondary vertices and the minimum distance of approach of charged particle trajectories to the primary interaction at D0~\cite{d0btag}. 

Events are selected at CDF with dedicated online event selections (triggers) requiring at least two jets and using $b$-tagging information, while at least three jets are required at D0. For most of the data collected at D0, $b$-tagging information is also used in the trigger decision. Events are selected offline with at least three jets within the fiducial region. This is defined by requirements on the momentum transverse to the proton beam direction, $p_T$, and detector pseudorapidity, $\eta_{\rm det} = -\ln (\tan(\theta /2))$, of jets, where $\theta$ is the polar angle with respect to the proton beam direction and an origin at the center of the detector: three jets with $p_T > 20$~GeV and $|\etadet| < 2.0$ at CDF and three jets with $p_T > 20$~GeV and at least two jets with $p_T > 25$~GeV, $|\etadet| < 2.5$ at D0. Events with more jets are accepted at CDF but only the leading (in $p_T$) four jets are used in the analysis. Exclusive channels with exactly three or exactly four jets are used at D0. The signal sample is defined by both experiments requiring at least three $b$-tagged jets. In the CDF inclusive and the D0 4-jet channel the two leading jets and at least one of the third and fourth jets are required to be $b$ tagged. In addition, a large sample requiring only two $b$-tagged jets is used by both experiments to build models of the multijet backgrounds.

At CDF, the background is modeled using a sum of templates of the invariant mass distribution of the leading two jets, representing contributions from different background modes categorized according to kinematics and flavor content.  The templates are constructed from events in the double-tagged data sample where at least one of the leading two jets is $b$-tagged. The events are then weighted according to the probability for at least one of the jets with no $b$ tag to pass the tagging requirements under three different assumptions as to whether it arises from a $b$ quark, a $c$ quark or a light quark (or gluon). This results in six different mass templates. The separation between the different components is enhanced by using an additional jet-flavor-sensitive discriminant, $x_{\rm tags}$, that makes use of kinematic properties of the charged particles from secondary vertices associated with each $b$-tagged jet.

In the D0 analysis the background model is formed by correcting the shape of the dijet invariant mass distribution of a data sample with two $b$-tagged jets using the ratio of simulated multijet samples with three $b$ tags and two $b$ tags. The simulated samples are generated using {\sc alpgen}~\cite{alpgen} with showering and hadronization carried out using {\sc pythia}~\cite{pythia} and detailed simulation of the detector using {\sc geant}~\cite{geant}. Their flavor composition, in terms of the relative numbers of $b$, $c$ and light jets per event, is determined from a simultaneous fit to the data across samples with differing numbers of tagged jets, different $b$-tagger operating points, and in small intervals of the scalar sum of the transverse momenta of the jets. The shape correction is applied as a function of the dijet invariant mass and the value of a likelihood-ratio discriminant, $\cal{L}$, designed to select signal-like events in preference to background-like events. Only the two possible jet pairings from the leading jet plus either the second or third leading jet are considered when forming Higgs candidates, and the choice that gives the highest value of $\cal{L} >$~0.65 is selected. If neither pairing in the event is above this threshold then the event is discarded. Systematic uncertainties are assessed to take into account the imperfect modeling of the background simulation. However, the definition of the correction as a ratio of distributions significantly reduces the sensitivity of the final model to these imperfections.
Small contributions to the background from top-quark-pair production are simulated using {\sc alpgen} and {\sc pythia}. Backgrounds from other sources, such as $Z\rightarrow b\bar{b}$ and single-top-quark production are negligible.

The efficiency for selecting signal is estimated for both analyses using events generated with {\sc pythia}. Associated production of a $\phi$ boson and a $b$ quark, $gb\rightarrow \phi b$, with subsequent decay $\phi\rightarrow b\bar{b}$, is used to model the signal. The signal cross section, experimental acceptance and the kinematics are corrected to next-to-leading order (NLO) using {\sc mcfm}~\cite{cite:MCFM,footnote1}, weighting events as a function of the kinematics of the leading $b$-quark jet not associated with the Higgs decay.

Approximately 11~500 events are selected by CDF with at least three $b$ tags and an estimated product of signal efficiency and acceptance varying as a function of $M_{\phi}$ between 0.18\% at $M_{\phi}=90$~GeV and 0.8\% at $M_{\phi}=200$~GeV. Approximately 15~000 and 11~000 events with at least three $b$-tagged jets are accepted in the D0 three-jet and four-jet channels, respectively. The corresponding signal efficiencies for a Higgs of mass $M_{\phi}=200$~GeV are 1.2\% and 0.6\%.
 
The channels are combined and exclusion limits set, using the modified frequentist technique~\cite{cls1,cls2}, with a log likelihood ratio (LLR) as test statistic: 
\begin{equation}
  \mathrm{LLR} = -2\ln\frac{p(X|H_1)}{p(X|H_0)},
\end{equation}
where $p$ represents the probabilities that the data, $X$, are drawn from the background-only ($H_0$), and signal-plus-background ($H_1$) hypotheses, respectively. The likelihoods are constructed from the binned two-dimensional distribution of the dijet invariant mass versus the $x_{\rm tags}$ discriminant for CDF and the binned one-dimensional dijet invariant mass distribution for D0.  Theoretical predictions of the absolute rates for the multijet backgrounds have large uncertainties. Therefore,  additional scale factors are applied to the background yield in the D0 analysis and the individual templates in the CDF analysis and introduced into the likelihood as parameters with no external constraints. Systematic uncertainties are introduced as either shape or normalization variations to the model probability density functions. These systematics are governed by nuisance parameters together with Gaussian constraint terms where appropriate~\cite{collie}. Sources of uncertainty related to a common component of the luminosity determination and the theoretical modeling of the signal production are considered fully correlated. All other sources of uncertainty are assumed to be uncorrelated. The modeling of the $b$-jet-tagging efficiency and the contamination from light-quark and gluon jets fake are the dominant sources of uncertainty on the background model. These are implemented as uncertainties that affect the shape of the distributions entering the likelihood. Additional sub-dominant uncertainties are considered in the D0 analysis arising in the modeling of the trigger, jet efficiency, jet energy scale and jet resolution. While most of those effects have a negligible impact on the background model, effects that are dependent on the differences between $b$, light, and gluon jets can be significant. Dominant experimental systematic uncertainties on the signal model can be attributed to luminosity~(6\%), $b$-tagging efficiency~(11-18\%) and jet energy scale (2-10\% depending on the $\phi$ boson mass hypothesis).

Limits on the product of the cross section and the branching ratio using the LLR test statistic are extracted. The limits are model-independent, apart from  assuming a single narrow Higgs boson mass peak, dominated by experimental resolution effects. These are summarized in Table \ref{tab:xsec}, and presented in Fig. \ref{fig:xsec}. The combination gives a sensitivity that is better than the D0 expected limit alone by $\approx25\%$ at $M_\phi=100$~GeV, steadily falling to $<1\%$ by $M_{\phi}=300$~GeV. Excesses of events above the SM background expectation are observed for $M_{\phi}=120$ and 140~GeV with significances of 2.5 standard deviations and 2.6 standard deviations, respectively. These are driven by the excesses observed in the individual contributing analyses of 2.8 standard deviations at $M_{\phi}=150$~GeV at CDF and 2.5 standard deviations at $M_{\phi}=120$~GeV at D0. A standard convention~\cite{trialsfactor} is used to account for the effect that it is more likely to find a deviation (under the background-only) hypothesis when several mass regions are probed compared with only a single hypothesis. The significance of the excesses in the combined analysis is reduced to $\approx$~2 standard deviations. 

\begin{table}
\centering
\begin{tabular}{c|c|ccccc}
\hline \hline
$M_{\phi}$  &  Obs. & \multicolumn{5}{c}{Expected (pb)} \\
(GeV)    &  (pb) & -2 s.d. & -1 s.d. & median & +1 s.d. & +2 s.d. \\ \hline
90  &38   &30   & 41   & 57   & 81  & 110 \\
100 &43   &22   & 30   & 42   & 62  & 86  \\
110 &43   &15   & 20   & 27   & 39  & 53  \\
120 &44   &12   & 15   & 20   & 29  & 39  \\
130 &25   &8.1  & 9.6  & 14   & 19  & 25  \\
140 &26   &5.1  & 7.3  & 10   & 14  & 20  \\
150 &18   &4.1  & 5.5  & 7.4  & 11  & 15  \\
160 &12   &3.3  & 4.5  & 6.0  & 8.5 & 12  \\
170 &9.4  &2.5  & 3.6  & 5.0  & 7.0 & 9.5 \\
180 &7.1  &2.4  & 3.2  & 4.2  & 6.0 & 8.2 \\
190 &5.7  &2.2  & 2.5  & 4.0  & 5.0 & 7.0 \\
200 &5.0  &2.0  & 2.4  & 3.0  & 4.4 & 6.0 \\
210 &3.8  &1.6  & 2.2  & 2.6  & 3.7 & 5.0 \\
220 &3.3  &1.2  & 1.7  & 2.2  & 3.2 & 4.7 \\
230 &2.5  &1.0  & 1.5  & 2.0  & 3.0 & 4.1 \\
240 &2.0  &0.9 & 1.2  & 1.8  & 2.5 & 3.5 \\
250 &2.0  &0.9 & 1.1  & 1.6  & 2.3 & 3.2 \\
260 &1.7  &0.7 & 1.0 & 1.4  & 2.2 & 2.8 \\
270 &1.3  &0.7 & 0.9 & 1.2  & 2.0 & 2.4 \\
280 &1.1  &0.6 & 0.8 & 1.1  & 1.7 & 2.4 \\
290 &0.82 &0.6 & 0.8 & 1.1  & 1.9 & 2.4 \\
300 &0.71 &0.5 & 0.7 & 1.0 & 1.6 & 2.3 \\ \hline \hline
\end{tabular}
\caption{Observed and expected upper limits at the 95\% C.L. on the product of cross section and branching ratio 
$\sigma(gb\rightarrow\phi b)\times BR(\phi\rightarrow bb)$,
within the acceptance for the highest $p_T$ $b$ quark not arising from the Higgs decay~\cite{footnote1}. Expected limits are given for the median and for $\pm$1 and $\pm$2 standard deviation (s.d.) variations of the background expectation.}
\label{tab:xsec}
\end{table}

\begin{figure}[!htbp]
\centering
\includegraphics[width=0.95\linewidth]{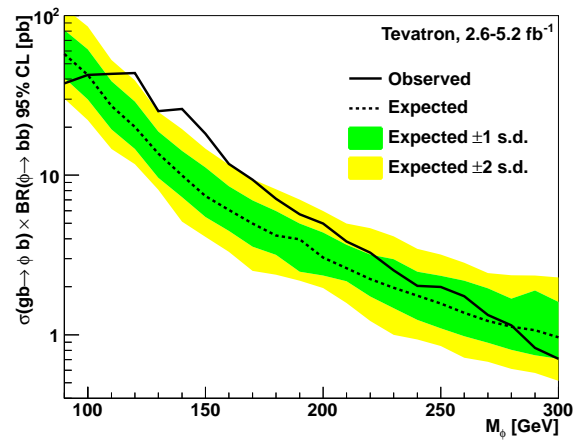}
\caption{Model independent 95\% C.L. upper limits on the product of cross section and branching ratio for the combined analyses, assuming a mass degeneracy between two of the three neutral bosons and a Higgs boson width significantly smaller than the experimental resolution. The dark and light shaded regions (color online) correspond to the one and two standard deviation bands around the median expected limit.}
\label{fig:xsec}
\end{figure}

Though these limits are the key results of this search, it is interesting to interpret them in terms of constraints on benchmark models within the MSSM.  As a consequence of the enhanced couplings to $b$ quarks at large $\tan\beta$, the total width of the Higgs boson increases with $\tan\beta$. When this width becomes comparable to the experimental resolution of 15-20\% there is an impact on the sensitivity of the search. When interpreting the exclusion within the MSSM, the width of the Higgs boson and the enhancement of the product of cross section and branching ratio above that of the SM are calculated using {\sc feynhiggs}~\cite{feynhiggs1,feynhiggs2,feynhiggs3,feynhiggs4,deltab1,deltab2}. The width is included in the simulation of the signal as a function of mass and $\tan\beta$ by convoluting a relativistic Breit-Wigner function with the NLO cross section~\cite{cite:MCFM}. Additional uncertainties for these model-specific limits are considered that otherwise cancel in the model-independent limit. For comparison these are derived as in previous results~\cite{hbbrun2a, cite:D0bbb, cite:CDFbbb}: uncertainties on the SM signal cross section are derived from varying the factorization and renormalization scales by a factor of two and from uncertainties on the parton distribution functions. The uncertainties assessed from scale variation are taken to be~10\% and the parton distribution functions contribute an additional 3.5-13\%, depending on the mass hypothesis.

The masses and couplings of the Higgs bosons in the MSSM depend, in addition to $\tan\beta$ and $M_A$, on other parameters through radiative corrections. Limits on $\tan\beta$ as a function of $M_A$ are derived for the $m^{\rm max}_{h}$ scenario~\cite{mhmax} that favors the $b\bar{b}$ final state,  assuming a CP-conserving Higgs sector~\cite{Carena:2005ek} and a negative value of the Higgs sector bilinear coupling $\mu$.  Figure~\ref{fig:mssm} shows exclusion limits in the $(\tanb,M_{A})$ plane for this scenario. Adding a further potential theoretical uncertainty on the signal cross section of 20\%, independent of $M_{\phi}$, would lead to an increase of 5\% in the \tanb\ limit.

\begin{figure}[!htb]
\centering
\includegraphics[width=0.95\linewidth]{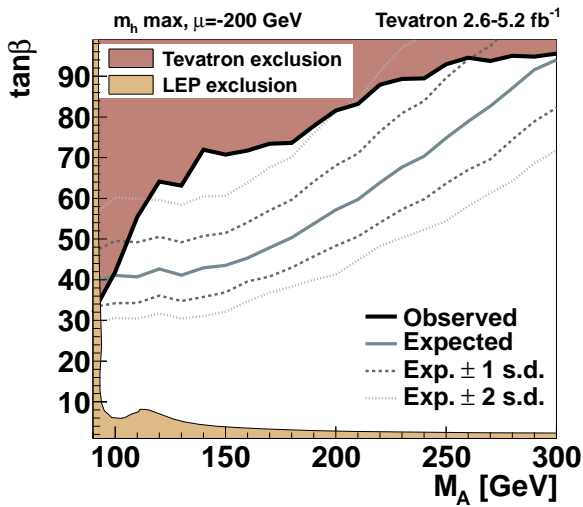}
\caption{95\% C.L. lower limit in the ($M_A$, tan$\beta$) plane for the $m^{\rm max}_h, \mu = -200$~GeV, including Higgs boson width effects. The exclusion limit obtained from the LEP experiments is also shown.}
\label{fig:mssm}
\end{figure}

In summary, the combination of results on neutral Higgs boson searches in multi-$b$-jet events from CDF and D0 has been presented. Upper limits are set on the product of the cross section and branching ratio and constraints are placed in the $(\tanb,M_{A})$ plane for a particular MSSM scenario. This combination, with more than half the integrated luminosity from the Tevatron still to be analyzed, provides the most stringent limits on neutral Higgs boson production and decay in the multi-$b$-jet mode.

%
We thank the Fermilab staff and technical staffs of the participating institutions for their vital contributions and acknowledge support from the
DOE and NSF (USA),
ARC (Australia),
CNPq, FAPERJ, FAPESP and FUNDUNESP (Brazil),
NSERC (Canada),
NSC, CAS and CNSF (China),
Colciencias (Colombia),
MSMT and GACR (Czech Republic),
the Academy of Finland,
CEA and CNRS/IN2P3 (France),
BMBF and DFG (Germany),
DAE and DST (India),
SFI (Ireland),
INFN (Italy),
MEXT (Japan),
the Korean World Class University Program and NRF (Korea),
CONACyT (Mexico),
FOM (Netherlands),
MON, NRC KI and RFBR (Russia),
the Slovak R\&D Agency, 
the Ministerio de Ciencia e Innovaci\'{o}n, and Programa Consolider-Ingenio 2010 (Spain),
The Swedish Research Council (Sweden),
SNSF (Switzerland),
STFC and the Royal Society (United Kingdom),
and the A.P Sloan Foundation (USA).


\begin{thebibliography}{99}
\bibitem{higgs1} F.~Englert and R.~Brout, Phys. Rev. Lett. {\bf 13}, 321 (1964).
\bibitem{higgs2} P.W.~Higgs, Phys. Lett. {\bf 12}, 132 (1964).
\bibitem{higgs3} P.W.~Higgs, Phys. Rev. Lett. {\bf 13}, 508 (1964).
\bibitem{higgs4} G.S.~Guralnik, C.R.~Hagen, and T.W.B.~Kibble, Phys. Rev. Lett. {\bf 13}, 585 (1964).
\bibitem{higgs5} P.W.~Higgs, Phys. Rev. {\bf 145}, 1156 (1966).
\bibitem{higgs6} T.W.B.~Kibble, Phys. Rev. {\bf 155}, 1554 (1967).
\bibitem{2HDM} V.~Barger, J.L~Hewett, and R.J.N.~Phillips, Phys. Rev. D {\bf 41}, 3421 (1990).
\bibitem{mssm1} H.P.~Nilles, Phys. Rep. {\bf 110}, 1 (1984). 
\bibitem{mssm2} H.E.~Haber and G.L.~Kane, Phys. Rep. {\bf 117}, 75 (1985). 
\bibitem{topbottom} B. Ananthanarayan, G. Lazarides, and Q. Shafi, Phys. Rev. D {\bf 44}, 1613 (1991).

\bibitem{darkmatter} V. Barger and C. Kao,  Phys. Lett. B {\bf 518}, 117 (2001).

\bibitem{octet1} B.A.~Dobrescu, K.~Kong, and R.~Mahbubani, Phys. Lett. B {\bf 670}, 119 (2008).
\bibitem{octet2} M.~Gerbush, T.J.~Khoo, D.J.~Phalen, A.~Pierce, and D.~Tucker-Smith, Phys. Rev. D {\bf 77}, 095003 (2008).
\bibitem{octet3} Y.~Bai and B.A.~Dobrescu,  J.\ High Energy Phys.\ 07, 100 (2011).
\bibitem{cite:LEP_exclu} The ALEPH Collaboration, The DELPHI Collaboration, The L3 Collaboration, and The OPAL Collaboration, Eur. Phys. J. C {\bf 47}, 547 (2006).

\bibitem{cite:CDF_exclu1} T.~Affolder {\sl et al.} (CDF Collaboration), \jprl{86}, 4472 (2001).
\bibitem{cite:CDF_exclu2} A.~Abulencia {\sl et al.} (CDF Collaboration), \jprl{96}, 011802 (2006).
\bibitem{cite:CDFbbb} T.~Aaltonen {\sl et al.} (CDF Collaboration), Phys. Rev. D {\bf 85}, 032005 (2012).
\bibitem{cite:D0_exclu} V.M.~Abazov {\sl et al.} (D0 Collaboration), \jprl{95}, 151801 (2005).

\bibitem{tautau1} V.M.~Abazov {\sl et al.} (D0 Collaboration), \jprl{97}, 121802 (2006).
\bibitem{tautau2}  V.M.~Abazov {\sl et al.} (D0 Collaboration), \jprl{101}, 071804 (2008).

\bibitem{hbbrun2a} V.M.~Abazov {\sl et al.} (D0 Collaboration), \jprl{101}, 221802 (2008).
\bibitem{tautau3}  V.M.~Abazov {\sl et al.} (D0 Collaboration), \jprl{102}, 051804 (2009).
\bibitem{tautau4}  V.M.~Abazov {\sl et al.} (D0 Collaboration), \jprl{104}, 151801 (2010).
\bibitem{cite:D0bbb} V.M.~Abazov {\sl et al.} (D0 Collaboration), Phys. Lett. B {\bf 698}, 97 (2011).
\bibitem{tautau5} V.M.~Abazov {\sl et al.} (D0 Collaboration), \jprl{107}, 121801 (2011).
\bibitem{tautau6} V.M.~Abazov {\sl et al.} (D0 Collaboration), Phys. Lett. B {\bf 707}, 323 (2012).
\bibitem{d0mssm} V.M.~Abazov {\sl et al.} (D0 Collaboration), Phys. Lett. B {\bf 710}, 569 (2012).

\bibitem{cite:CMStautau1}  CMS Collaboration, \jprl{106}, 231801 (2011).
\bibitem{cite:ATLAStautau} ATLAS Collaboration, Phys. Lett. B {\bf 705}, 174 (2011).
\bibitem{cite:CMStautau2} CMS Collaboration, Phys. Lett. B {\bf 713}, 68 (2012).
\bibitem{cdfdet} D.~Acosta {\sl et al.} (CDF Collaboration), Phys. Rev. D {\bf 71} 032001 (2005).
\bibitem{run2det1}
V.M.~Abazov {\sl et al.} (D0 Collaboration), Nucl. Instrum. Methods Phys. Res. A {\bf 565}, 463  (2006).
\bibitem{run2det2} M.~Abolins {\sl et al.} Nucl. Instrum. Methods Phys. Res. A {\bf 584}, 75 (2008).
\bibitem{run2det3} R.~Angstadt {\sl et al.} Nucl. Instrum. Methods Phys. Res. A {\bf 622}, 298 (2010).
\bibitem{cone}
  G.~Blazey {\it et al.}, arXiv:hep-ex/0005012 (2000).
  
\bibitem{cdfbtag} D.~Acosta {\sl et al.} (CDF Collaboration), Phys. Rev. D {\bf 71}, 052003 (2005).
\bibitem{d0btag} V.M.~Abazov {\sl et al.} (D0 Collaboration), Nucl. Instrum. Methods in Phys. Res. Sect. A {\bf 620}, 490 (2010).
  \bibitem{alpgen}
    M.L. Mangano {\sl et al.}, J.\ High Energy Phys.\  07, 001 (2003). {\sc alpgen} version 2.11 is used.

  \bibitem{pythia}
  T. Sj\"{o}strand, S. Mrenna, and P. Skands, J. High Energy Phys. 05, 026 (2006). Version 6.409 is used.
  \bibitem{geant}
   R. Brun and F. Carminati, CERN program library long writeup W5013 (1993).

  \bibitem{cite:MCFM}    J. Campbell, R.K. Ellis, F. Maltoni, and S. Willenbrock,  Phys.\ Rev.\ D~{\bf 67}, 095002 (2003).
 
\bibitem{footnote1}  Cross sections are quoted with respect to the requirements that the highest-$p_T$ $b$ quark not from the Higgs decay has $p_T > 12$~GeV and $|\eta| < 5.0$.
\bibitem{cls1}
    T. Junk, Nucl. Instrum. Methods Phys. Res. A {\bf 434}, 435 (1999).
\bibitem{cls2} A. Read, Nucl. Instrum. Methods Phys. Res. A {\bf 425}, 357 (1999).
\bibitem{collie} W. Fisher, FERMILAB-TM-2386-E (2007).
\bibitem{trialsfactor} The ALEPH Collaboration, The DELPHI Collaboration, The L3 Collaboration, and The OPAL Collaboration, Phys. Lett. B {\bf 565}, 61 (2003).
\bibitem{feynhiggs1} S.~Heinemeyer, W.~Hollik, and G.~Weiglein, Eur.\ Phys.\ J.\  C {\bf 9}, 343 (1999). {\sc feynhiggs} version 2.6.8 is used.
\bibitem{feynhiggs2} S.~Heinemeyer, W.~Hollik, and G.~Weiglein, Comput. Phys. Commun. {\bf 124}, 76 (2000).
\bibitem{feynhiggs3} G.~Degrassi, S.~Heinemeyer, W.~Hollik, P.~Slavich, and G.~Weiglein, Eur.\ Phys.\ J.\  C {\bf 28}, 133 (2003).
\bibitem{feynhiggs4} M.~Frank, T.~Hahn, S.~Heinemeyer, W.~Hollik, H.~Rzehak, and G.~Weiglein, J.\ High Energy Phys.\  02, 047 (2007). 
\bibitem{deltab1} L.~Hofer, U.~Nierste, and D.~Shere, J.\ High Energy Phys.\ 10, 081 (2009).
\bibitem{deltab2} D.~Noth and M.~Spira, Phys. Rev. Lett. {\bf 101}, 181801 (2008). 

\bibitem{mhmax}
$M_{\rm SUSY} = 1$ TeV, $X_{t} = 2$ TeV, $M_2 = 0.2$ TeV $|\mu| = 0.2$ TeV, and $m_g = 0.8$ TeV. 

\bibitem{Carena:2005ek}
  M.~Carena, S.~Heinemeyer, C.~E.~M.~Wagner, and G.~Weiglein,
  Eur.\ Phys.\ J.\  C {\bf 45}, 797 (2006).


\end{thebibliography}
\end{document}